\documentclass[peerreview,onecolumn,11pt,draftclsnofoot]{IEEEtran}
\usepackage{amsmath}
\usepackage{multirow}
\usepackage{amsfonts}
\usepackage{epsfig}
\usepackage{amssymb}
\usepackage{graphicx}
\usepackage{amssymb,amsmath}
\usepackage{cite}
\usepackage{color,soul}
\usepackage{amsmath}

\begin{document}
\title{Robust Additively Coupled Games}

\author{Saeedeh Parsaeefard, \IEEEmembership{Student Member, IEEE,}
        Ahmad R. Sharafat, \IEEEmembership{Senior Member, IEEE,}
and Mihaela van der Schaar, \IEEEmembership{Fellow, IEEE}
%
}
\maketitle

\begin{abstract}
We study the robust Nash equilibrium (RNE) for a class of games in communications systems and networks where the impact of users on each other is an additive function of their strategies. Each user measures this impact, which may be corrupted by uncertainty in feedback delays, estimation errors, movements of users, etc. To study the outcome of the game in which such uncertainties are encountered, we utilize the worst-case robust optimization theory. The existence and uniqueness conditions of RNE are derived using finite-dimensions variational inequalities. To describe the effect of uncertainty on the performance of the system, we use two criteria measured at the RNE and at the equilibrium of the game without uncertainty. The first is the difference between the respective social utility of users and, the second is the differences between the strategies of users at their respective equilibria. These differences are obtained for the case of a unique NE and multiple NEs. To reach the RNE, we propose a distributed algorithm based on the proximal response map and derive the conditions for its convergence. Simulations of the power control game in interference channels, and Jackson networks validate our analysis.
\end{abstract}


\section{Introduction}
\IEEEPARstart{D}{istributed} designs for multi-user communications networks and systems have been extensively used during the past decade to implement low-cost, scalable, and limited-message-passing networks. In doing so, transmitter and receiver pairs with local information determine their transmission strategies in an autonomous manner. To deploy such designs, it is essential to know whether they converge to a (preferably unique) equilibrium, and evaluate their performances at the emerging equilibrium/equilibria.

Strategic non-cooperative game theory provides an appropriate framework for analyzing and designing such environments where users (i.e., transmitter-receiver pairs) are rational and self-interested players that aim to maximize their own utilities by choosing their transmission strategies. The notion of Nash equilibrium (NE), at which no user can attain a higher utility by unilaterally changing its strategy, is frequently used to analyze the equilibrium point of non-cooperative games. To derive the conditions for NE's existence and uniqueness, different approaches such as fixed point theory, contraction mapping and \textit{variational inequalities (VI)} \cite{NEexistence,PangVI,Palomar2010} have been widely applied in both wired and wireless communication networks, including applications to flow and congestion control, network routing, and power control in interference channels \cite{ref1,ref2,ref3,mihaelastructure,Goldsmith1,Yates,Nash1,Nash2}.

However, there are numerous sources of uncertainty in measured parameter values of communication systems and networks such as joining or leaving new users, delays in the feedback channel, estimation errors and channel variations. Therefore, obtaining accurate values of users' interactions may not be practical, and considering uncertainty and proposing a robust approach are essential in designing reliable communications systems and networks.

To make a NE robust against uncertainties, two distinct approaches have been proposed in the literature:  the Bayesian approach where the statistics of uncertain parameters are considered and the utility of each user is probabilistically guaranteed, and the worst-case approach where a deterministic closed region, called the uncertainty region, is considered for the distance between the exact and the estimated values of uncertain parameters, and the utility of each user is guaranteed for any realization of uncertainty within the uncertainty region \cite{selecectedrobust,Robustscutari,Robusthaykin,Gershman}.

Both of these approaches have been applied to the power allocation problem in spectrum sharing environments and cognitive radio networks \cite{ProbabilisticIWFA,Robustnew,Robusthaykin,Robustscutari} to study the conditions for NE's existence and uniqueness, where the uncertain parameters are interference levels and channel gains. However, to incorporate robustness in communications systems and networks, there exist multiple challenges such as:  1) How to implement robustness in a wider class of problems in communication systems? 2) How to derive the conditions for existence and uniqueness of the robust NE (RNE)? 3) What is the impact of considering uncertainties on the system's performance at its equilibrium  compared to that of the case with no uncertainty? 4) How to design a distributed algorithm for reaching the robust equilibrium?

In this paper, we aim to answer the above questions using the worst-case robust optimization. In doing so, we consider a general class of games where the impact of users on each other is an additive function of their actions, which causes couplings between users. We refer to this class of game as the additively coupled games (ACGs). In the ACG, we consider that the users' observations of such impacts are uncertain due to variations in system parameters and changes in other users' strategies. Via the worst-case approach, we assume that uncertain observations by each user are bounded in the uncertainty region, and each user aims to maximize its utility for the worst-case condition of error. We refer to an ACG that considers uncertainty as a robust ACG (RACG), and an ACG that does not consider uncertainty as the nominal ACG (NACG). To study the conditions for existence and uniqueness of the RNE, we apply  \textit{VI} \cite{PangVI}, and show that with bounded and convex uncertainty, the RNE always exists. We also show that the RNE is a perturbed solution of \textit{VI}, and derive the condition for RNE's uniqueness based on the condition for NE's uniqueness.

Furthermore, we compare the performance of the system at the RNE with that at the NE in terms of two measures: 1) the difference between the users' strategies at the RNE and the NE, 2) the difference between the social utility at the RNE and at the NE. When the RNE is unique, we derive the upper bound for the difference between social utilities at the RNE and at the NE, and show that the social utility at the RNE is always less than that at the NE. However, obtaining these two measures is not straightforward when the NE is not unique. In this case, we demonstrate a condition in which the social utility at a RNE is higher than that at the corresponding NE. Finally, we apply the proximal response map to propose a distributed algorithm for reaching the RNE, and derive the conditions for its convergence.

The rest of this paper is organized as follows. In Section II, we summarize the system model of the NACG. In Section III, we introduce the RACG and its RNE. Section IV covers the existence and uniqueness conditions if the RNE. In Section V, we show that when the utility function is logarithmic, the RNE can be obtained via affine \textit{VI} (\textit{AVI}), and its uniqueness condition is simplified. In Section VI, we propose distributed algorithm for reaching RNE. In Section VII, we discuss the effects of robustness for the case of multiple Nash equilibria, followed by Section VIII, where we provide simulation results to illustrate our analytical developments for the power allocation problem and for the Jackson networks. Finally, conclusions are drawn in Section IX.

\section{System Model}
Consider a set of communication resources divided into $K$ orthogonal dimensions, e.g., frequency bands, time slots, and routes, which are shared between a set of users denoted by $\mathcal{N}=\{1,\cdots,N\}$, where each user consists of a transmitter and a receiver. We assume that users do not cooperate with each other, and formulate the resource allocation problem as a strategic non-cooperative game $\mathcal{G}=\{\mathcal{N},(v_n)_{ n \in\mathcal{N} },\mathcal{A}\}$, where $\mathcal{N}$ is the set of players (users) in the game, $\mathcal{A}=\prod_{n \in \mathcal{N}}\mathcal{A}_{n}$ is the joint strategy space of the game, and $\mathcal{A}_{n}\subseteq \mathbb{R}^K$ is the strategy space of user $n$ in which the strategy of each user is limited in each dimension. The sum of strategies of each user over all dimensions is bounded, i.e.,
\begin{equation}\label{An}
   \mathcal{A}_n=\{\textbf{a}_n=(a_{n}^{1},\cdots,a_{n}^{K})|a_{n}^{k} \in [a_{n,k}^{\texttt{min}},a_{n,k}^{\texttt{max}}]\,\qquad \text{and} \qquad\,\sum_{k=1}^{K}a_{n}^{k}\leq
  a_{n}^{\texttt{max}}\}
\end{equation}
where $a_{n,k}^{\texttt{min}}$ and $a_{n,k}^{\texttt{max}}$ is the minimum and the maximum transmission strategy of each user in each dimension and  $a_{n}^{\texttt{max}}$ is the bound on the sum of strategies of user $n$ over all dimensions, e.g., the maximum transmit power of each user. The function $v_{n}(\textbf{a}): \mathcal{A} \rightarrow \mathbb{R}$ is the utility function of user $n$ and depends on the chosen strategy vector of all users $\textbf{a}=[\textbf{a}_1,\cdots,\textbf{a}_{N}]$, where $\textbf{a}_{n} \in \mathcal{A}_n $ is the action of user $n$. The vector of actions of all users except user $n$ is denoted by $\textbf{a}_{-n} \in \mathcal{A}_{-n}$  where $\mathcal{A}_{-n}=\prod_{m \in \mathcal{N}, m\neq n}\mathcal{A}_{m}$ is the strategy space of all users except user $n$.

In a non-cooperative strategic game, each user aims to maximize its own utility subject to its strategy space as
\begin{equation}\label{1_opteachuser}
    \max_{\textbf{a}_{n} \in \mathcal{A}_{n}} v_{n}(\textbf{a}), \quad \forall n \in \mathcal{N}.
\end{equation}
which is the optimization problem of each user in the game. Let us assume that the utility of each user is a function of its action and the impact of the others users' actions over all dimensions, i.e.,
\begin{equation}\label{utilityseprale}
    v_n(\textbf{a}_n, \textbf{f}_n(\textbf{a}_{-n}, \textbf{x}_n))=\sum_{k=1}^{K} v_n^k(a_{n}^{k},f_{n}^{k}(\textbf{a}_{-n}, \textbf{x}_n))
\end{equation}
where $\textbf{f}_n(\textbf{a}_{-n}, \textbf{x}_{n})= [f_{n}^{1}(\textbf{a}_{-n}, \textbf{x}_n),...,f_{n}^{K}(\textbf{a}_{-n}, \textbf{x}_n)]$ is the $1 \times K$ vector of the additive impact of other users on user $n$ with the following elements
  \begin{equation}\label{}
        f_{n}^{k}(\textbf{a}_{-n}, \textbf{x}_n)=\sum_{m \in \mathcal{N}, m \neq n} a_{m}^{k}x_{nm}^{k}+y_{n}^{k}, \quad \forall k,
\end{equation}
where $\textbf{x}_n=[\textbf{x}_{n1},\cdots,\textbf{x}_{n(n-1)},\textbf{x}_{n(n+1)}, \cdots,\textbf{x}_{nN}, \textbf{y}_n]$ is the vector of system's parameters for user $n$, in which $\textbf{x}_{nm}=[x_{nm}^1, \cdots,x_{nm}^K]$, and $x_{nm}^k$ represents the system's parameters between user $n$ and user $m$ in dimension $k$, e.g., the channel gain between user $n$ and user $m$ in sub-channel $k$; $\textbf{y}_n=[y_n^1,\cdots,y_n^K]$ is the vector of the effect of system on user $n$, e.g., noise in sub-channels for user $n$, and $y_n^k$ is related to the $k^{\text{th}}$ dimension. Assuming $\textbf{f}_{n}^{k}(\textbf{a}_{-n}, \textbf{x}_n)$ as a linear function is very practical in communication networks. For example, the interference in each receiver is a linear and additive function of the actions of other users and channel gains plus noise. We also assume that \textbf{A1}) The utility function of each user is an increasing, twice differentiable, and concave function with respect to $\textbf{a}_n$ and has bounded gradients; \textbf{A2}) The utility function of each user is a decreasing, twice differentiable, and convex function with respect to $\textbf{f}_{n}(\textbf{a}_{-n}, \textbf{x}_n)$; \textbf{A3}) The utility function of each user is twice differentiable over $\textbf{a}_n$, and $\textbf{f}_n(\textbf{a}_{-n}, \textbf{x}_n)$. Note that \textbf{A1} is a typical utility function in communications networks \cite{CP}; \textbf{A2} is a practical assumption for sharing resources between non-cooperative users \cite{mihaelastructure}; and \textbf{A3} indicates that the utility function is continuously differentiable with respect to $\textbf{a}_n$ and $\textbf{f}_n(\textbf{a}_{-n}, \textbf{x}_n)$. As such, the above assumptions \textbf{A1}-\textbf{A3} are typical in communications systems and networks.

Interactions between users are studied at the NE, which corresponds to the strategy profile
$\textbf{a}^{*}=(\textbf{a}_{1}^{*},\cdots,\textbf{a}^{*}_{N})$ such that for any other strategy profile, we have  \cite{NEexistence}
\begin{equation}\label{NE}
    v_n(\textbf{a}^*_{n},\textbf{f}_n(\textbf{a}_{-n}^*, \textbf{x}_n))\geq v_n(\textbf{a}_{n},\textbf{f}_n(\textbf{a}_{-n}^*, \textbf{x}_n)), \quad n \in
\mathcal{N}.
\end{equation}
In what follows, we denote the achieved utility of user $n$ at the NE by $v_n^*$ and the social utility at the NE by $v^*=\sum_{n=1}^{N}v_n^*$, derive the conditions for NE's existence and uniqueness conditions via reformulating NE through \textit{VI}, and show that in this way, the NE and the RNE can be analyzed similarly.

\textbf{Remark 1}. Consider $\mathcal{F}(\textbf{a})=(\mathcal{F}_{n}(\textbf{a}))_{n=1}^{N}$, where
\begin{equation}\label{mappingG}
   \mathcal{F}_n(\textbf{a})=(- \nabla_{\textbf{a}_{n}} v_n(\textbf{a}_{n}, \textbf{f}_n(\textbf{a}_{-n}, \textbf{x}_n)),
\end{equation}
where $\nabla_{\textbf{a}_{n}} v_n(\textbf{a}_{n}, \textbf{f}_n(\textbf{a}_{-n}, \textbf{x}_n))$ denotes the column gradient vector of $ v_n(\textbf{a}_{n}, \textbf{f}_n(\textbf{a}_{-n}, \textbf{x}_n))$ with respect to $\textbf{a}_n$. The NE of $\mathcal{G}$ can be obtained by solving $VI(\mathcal{A}, \mathcal{F})$ (Proposition 1.4.2 in \cite{PangVI}) as
\begin{equation}\label{VImapNE}
    (\textbf{a}-\textbf{a}^*)\mathcal{F}(\textbf{a}^*) \geq 0, \forall \textbf{a} \in \mathcal{A},
\end{equation}
Since $v_{n}(\textbf{a}_{n},\textbf{f}_n(\textbf{a}_{-n}, \textbf{x}_n))$ is a continuous and concave function with respect to $\textbf{a}_{n}\in\mathcal{A}_n$, $\mathcal{F}(\textbf{a})$ is a continuous map. From (\ref{An}), the set $\mathcal{A}$ is convex and compact. Therefore, the solution set of $VI(\mathcal{A}, \mathcal{F})$ is nonempty and compact (Theorem 2.2.1 in \cite{PangVI}). Consequently, the NE of ACG exists.

\textbf{Remark 2}. Let us consider the following definitions of mapping $\mathcal{F}(\textbf{a})$,
\begin{eqnarray}
  \alpha_{n}(\textbf{a})\triangleq \text{smallest eigenvalue of}
     - \nabla^{2}_{\textbf{a}_{n}}v_{n}(\textbf{a}_{n},\textbf{f}_n(\textbf{a}_{-n}, \textbf{x}_n)) &\Longrightarrow& \alpha_{n}^{\text{min}}\triangleq \inf_{\textbf{a} \in \mathcal{A}}
    \alpha_{n}(\textbf{a}), \,\, \forall n \in \mathcal{N} \\
  \beta_{nm}(\textbf{a})\triangleq\|- \nabla_{\textbf{a}_{n}\textbf{a}_{m}} v_{n}(\textbf{a}_{n},\textbf{f}_n(\textbf{a}_{-n}, \textbf{x}_n))\|,
    \quad \forall n\neq m &\Longrightarrow& \beta_{nm}^{\text{max}}\triangleq \sup_{\textbf{a} \in \mathcal{A}}
    \beta_{n}(\textbf{a}), \,\, \forall n \in \mathcal{N}
 \end{eqnarray}
where $\nabla^{2}_{\textbf{a}_{n}} v_{n}(\textbf{a}_{n},\textbf{f}_n(\textbf{a}_{-n}, \textbf{x}_n))$ and $\nabla_{\textbf{a}_{n}\textbf{a}_{m}} v_{n}(\textbf{a}_{n},\textbf{f}_n(\textbf{a}_{-n}, \textbf{x}_n))$ are the $K \times K$ Jacobian matrices of $\mathcal{F}_n(\textbf{a})$ with respect to $\textbf{a}_n$ and $\textbf{a}_m$, respectively. Now, consider the $N \times N$ matrix $\boldsymbol{\Upsilon}$ whose elements are
\begin{eqnarray}
   [ \Upsilon]_{nm}= \left\{\begin{array}{c}
  \alpha_{n}^{\text{min}} \,\,\,\,\,\qquad\text{if} \qquad\, m=n\nonumber \\
-\beta_{nm}^{\text{max}}  \,\qquad  \text{if} \qquad\, m\neq n.\nonumber \\
\end{array} \right.
\end{eqnarray}
When $\boldsymbol{\Upsilon}$ is a $P$-matrix, the mapping
$\mathcal{F}(\textbf{a})$ is strictly monotone and the NE is unique (Theorem 12.5 in \cite{Palomar2010}).

This setup is applicable to some important problems in communications systems and networks, such as the additively coupled sum constrained game \cite{mihaelastructure}, which can be used to formulate, e.g., power control in interference channels, and the Jackson network. These two systems are chosen to illustrate the validity of our approach and analysis. Table \ref{tableexample} contains the parameters of the power control, and the Jackson network games.

The power control in interference channels can be formulated by a ACG, where each player, consisting of a transmitter and receiver pair, competes with other players to maximize its total data rate over $K$ distinct sub-channels. The strategy of each user is its transmit power in $K$ sub-channels, where $h_{nn}^{k}$ is the direct channel gain between the transmitter and receiver pair $n$, and $\bar{h}_{nm}^{k}=\frac{h_{nm}^{k}}{h_{nn}^{k}}$ and $\bar{\sigma}_{n}^{k}=\frac{\sigma_{n}^{k}}{h_{nn}^{k}}$ are the normalized interference channel gain between transmitter $m$ and receiver $n$, and the normalized noise power in sub-channel $k$, respectively. For the Jackson network in Table \ref{tableexample}, arriving packets to node $n$ are split into $\mathcal{K}=[1,\cdots,K]$ traffic classes, and the input rate and service rate for class $k$ are $\psi_{n}^{k}$ and $\mu_{n}^{k}$, respectively. Here, $\psi_{n}^{k}$ is the strategy of player $n$ in dimension $k$. The total rate is subject to the minimum rate constraint, i.e., $\sum_{k=1}^{K}\psi_{n}^{k}\geq \psi_{n}^\text{min}$. A packet of class $k$ completing service at node $m$ is routed to node $n$ with probability $r_{nm}^{k}$, or exit the network with probability $r_{m0}^{k}=1-\sum_{n=1}^{N}r_{nm}^{k}$. In this scenario, we denote $[R^{k}]_{nm}=r_{nm}^{k}$, $\boldsymbol{\Theta}^{k}=(1-\textbf{R}^{k})^{-1}$, and $\nu_{nm}^{k}=[\Theta^{k}]_{nm}$. It can be shown that the user's utility for minimizing $M/M/1$ queueing delay can be expressed by $d_{n}(\boldsymbol{\Psi})=\sum_{k=1}^{K}\frac{1}{\mu_{n}^{k}-\sum_{m=1}^{N} \nu_{nm}^{k}\psi_{m}^{k}}$, where $\boldsymbol{\Psi}=[\Psi_1,\cdots,\Psi_{N}]$, and $\Psi_n=[\psi_n^1,\cdots,\psi_n^K]$\;\;\;$\forall n \in \mathcal{N}$. The optimization problem can be rewritten by maximizing $\sum_{k=1}^{K}\mu_{n}^{k}-\sum_{m=1}^{N} \nu_{nm}^{k}\psi_{m}^{k}$ subject to the minimum data rate constraint for each user.

\section{Robust Games}

As stated earlier, users may encounter different sources of uncertainty caused by variations in $\textbf{a}_{-n}$ and/or  $\textbf{x}_{nm}$, which cause variations in the utility function of each user, and prevent users from attaining their expected performance. To deal with such issues, we assume that all uncertainties for a given user can be modeled by variations in the user's observation $\textbf{f}_{n}(\textbf{a}_{-n}, \textbf{x}_n)$, i.e.,
\begin{equation}\label{modelofucneratinty}
    \widetilde{\textbf{f}}_{n}(\textbf{a}_{-n}, \textbf{x}_n)=\textbf{f}_{n}(\textbf{a}_{-n}, \textbf{x}_n)+\widehat{\textbf{f}}_{n}(\textbf{a}_{-n}, \textbf{x}_n),
\end{equation}
where $\widetilde{\textbf{f}}_{n}(\textbf{a}_{-n}, \textbf{x}_n)=[\widetilde{f}^1_{n}(\textbf{a}_{-n}, \textbf{x}_n),\cdots,\widetilde{f}^K_{n}(\textbf{a}_{-n}, \textbf{x}_n)]$, $\textbf{f}_{n}(\textbf{a}_{-n}, \textbf{x}_n)=[f^1_{n}(\textbf{a}_{-n}, \textbf{x}_n),\cdots,f^K_{n}(\textbf{a}_{-n}, \textbf{x}_n)]$, and $\widehat{\textbf{f}}_{n}(\textbf{a}_{-n}, \textbf{x}_n)=[\widehat{f}^1_{n}(\textbf{a}_{-n}, \textbf{x}_n),\cdots,\widehat{f}^K_{n}(\textbf{a}_{-n}, \textbf{x}_n)]$ are the actual observation, the nominal observation, and the error in observation of user $n$, respectively. 
In the worst-case robust optimization theory, uncertainties are assumed to be bounded to the uncertainty region, stated by
 \begin{equation}\label{III-1}
\Re_{n}(\textbf{a}_{-n})=\{ \widetilde{\textbf{f}}_{n}(\textbf{a}_{-n}, \textbf{x}_n) \in \Re_{n}(\textbf{a}_{-n}) | \,\, \| \widehat{\textbf{f}}_{n}(\textbf{a}_{-n}, \textbf{x}_n) \|_p \leq \varepsilon_{n} \} \quad , \forall n  \in \mathcal{N}
 \end{equation}
where $\parallel \textbf{t} \parallel_p=\sqrt[p]{(\sum_{i=1}^{N}t_{i}^{p})}$ denotes the linear norm with order $p>1$. In communication and network systems, the ellipsoid region, i.e., $p=2$, has been commonly used to model uncertainty  \cite{Gershman,RobustCognitiveBeamforming,Centralized}. We also use the norm with $p=2$ in our robust game, and denote the uncertainty region by $\Re_{n}(\textbf{a}_{-n})$ so as to indicate that it is an additive function of the actions of other users and system parameters, i.e., it is not a fix region.

The effect of uncertainty in $\widetilde{\textbf{f}}_{n}(\textbf{a}_{-n}, \textbf{x}_n)$ is highlighted by a new variable in the utility function of each user as
\begin{equation}\label{utilityrobsut}
    u_{n}(\textbf{a}_n, \widetilde{\textbf{f}}_n(\textbf{a}_{-n}, \textbf{x}_n))= \sum_{k=1}^{K} u_{n}^{k}(a_{n}^{k}, \widetilde{f}_{n}^{k}(\textbf{a}_{-n}, \textbf{x}_n)) , \quad \forall n \in \mathcal{N}
\end{equation}
in such a way that
\begin{equation}\label{utilityrobust2}
    v_{n}(\textbf{a}_n, \textbf{f}_n(\textbf{a}_{-n}, \textbf{x}_n))=  u_{n}(\textbf{a}_n, \widetilde{\textbf{f}}_n(\textbf{a}_{-n}, \textbf{x}_n))|_{\varepsilon_{n}=0} , \quad \forall n \in \mathcal{N}
\end{equation}
The objective of the worst-case approach is to find the optimal strategy for each user that optimizes its utility under the worst condition of error in the uncertainty region. In this approach, from \textbf{A2}, the optimization problem of each user can be formulated as \cite{Robustgame}
\begin{equation}\label{optrobsut}
   \widetilde{u}_{n} = \max_{\textbf{a}_{n} \in \mathcal{A}_n}  \min_{\widetilde{\textbf{f}}_{n}(\textbf{a}_{-n}, \textbf{x}_n)\in \Re_{n}(\textbf{a}_{-n})} u_{n}(\textbf{a}_n, \widetilde{\textbf{f}}_n(\textbf{a}_{-n}, \textbf{x}_n)),
\end{equation}
where $\widetilde{u}_{n}$ is the achieved utility of user $n$ in the worst-case approach. The domain of optimization problem (\ref{optrobsut}) is defined by
\begin{equation}\label{Domain}
    \widehat{\mathcal{A}}_{n}(\textbf{a}_{-n})=\mathcal{A}_n \times \Re_{n}(\textbf{a}_{-n})
\end{equation}
which is a function of other users' strategy. We represent the RACG by
$\widetilde{\mathcal{G}}=\{\mathcal{N},(u_n)_{n\in \mathcal{N}},\widehat{\mathcal{A}}\}$ where $\widehat{\mathcal{A}}=\prod_{n=1}^{N}\widehat{\mathcal{A}}_{n}(\textbf{a}_{-n})$. The solution to (\ref{utilityrobsut}) for user $n$ is a pair $(\widetilde{\textbf{a}}'_n, \textbf{f}'_n(\textbf{a}_{-n}, \textbf{x}_n)) \in \mathcal{A}_n \times \Re_{n}(\textbf{a}_{-n})$ that satisfies \cite{Robustscutari}
\begin{equation}\label{utilityrobsutsaddlepoint}
  \max_{\textbf{a}_{n} \in \mathcal{A}_n}  u_n(\textbf{a}_{n},\textbf{f}'_n(\textbf{a}_{-n}, \textbf{x}_n))= u_n(\widetilde{\textbf{a}}'_{n},\textbf{f}'_n(\textbf{a}_{-n}, \textbf{x}_n))= \min_{\textbf{f}_n(\textbf{a}_{-n}, \textbf{x}_n) \in \Re_{n}(\textbf{a}_{-n})} u_n(\widetilde{\textbf{a}}'_{n},\textbf{f}_n)
\end{equation}
which is the saddle point of (\ref{utilityrobsut}). Using the above, the equilibrium of the robust game $\mathcal{G}$ is defined below.

\textbf{Definition 2.} The RNE of RACG corresponds to the strategy profile $\widetilde{\textbf{a}}^{*}=(\widetilde{\textbf{a}}_{1}^{*},\cdots,\widetilde{\textbf{a}}^{*}_{N})$
if and only if for any other strategy profile $\widetilde{\textbf{a}}_{n}$ we have
\cite{Robustgame}
\begin{equation}\label{RNEdefinition}
 \min_{\widetilde{\textbf{f}}_{n}(\textbf{a}_{-n}, \textbf{x}_n)\in \Re_{n}}
u_n(\widetilde{\textbf{a}}_{n}^*,\widetilde{\textbf{f}}_{n}(\textbf{a}^*_{-n}, \textbf{x}_n)) \geq \min_{\widetilde{\textbf{f}}_{n}(\textbf{a}_{-n}, \textbf{x}_n)\in \Re_{n}} u_n(\widetilde{\textbf{a}}_{n},\widetilde{\textbf{f}}_n(\textbf{a}_{-n}^*, \textbf{x}_n)) \quad\quad \forall \,\,
\widetilde{\textbf{a}}_{n} \in \mathcal{A}_n(\textbf{a}_{-n})
\end{equation}
We denote the achieved utility of user $n$ at the RNE by $\widetilde{u}_n^*$ and the social utility at the RNE by $\widetilde{u}^*=\sum_{n=1}^{N}\widetilde{u}_n^*$.

\section{RNE Analysis: Existence and Uniqueness Conditions}

Now we derive the characteristics of the RNE in the RACG from the NE in the NACG. For convenience, in what follows, we omit the arguments $\textbf{a}_{-n}$ and $\textbf{x}_n$ in $\widetilde{\textbf{f}}_n(\textbf{a}_{-n}, \textbf{x}_n)$.

\subsection{Existence of the RNE}

To analyze the existence of RNE, we encounter two problems. First, by considering uncertainty in the utility of each user, the utility may become non-convex, and analyzing RNE may become impossible. Second, the strategy space of user $n$ changes to $\widehat{\mathcal{A}}_{n}(\textbf{a}_{-n})=\mathcal{A}_n \times \Re_{n}(\textbf{a}_{-n})$ which is not a fix set and is a function of the other users' actions. Therefore, convexity of the optimization problem of each user is not a sufficient condition for the existence of RNE, meaning that we need to utilize \textit{VI} in the sequel.

\textbf{Lemma 1.} 1) For the uncertainty region in (\ref{III-1}), the strategy of each user is a convex, bounded, and closed set. 2) $\Psi_n(\textbf{a}_n, \textbf{a}_{-n})$ is a concave and continuous differentiable function of $\textbf{a}_n$ for every $\textbf{a}_{-n}$, where,
\begin{equation}\label{psi2}
    \Psi_n (\textbf{a}_n, \textbf{a}_{-n})= \min_{\widetilde{\textbf{f}}_{n}\in \Re_{n}(\textbf{a}_{-n})} u_{n}(\textbf{a}_n, \widetilde{\textbf{f}}_n) =u_n(\textbf{a}_n, \widetilde{\textbf{f}}^{*}_{n})
\end{equation}
and
\begin{equation}\label{followerstartegyspace2}
  \widetilde{\textbf{f}}^{*}_{n}=\textbf{f}_n- \varepsilon_n \boldsymbol{\vartheta}_{n}
\end{equation}
where $\widetilde{\textbf{f}}_n^*=[\widetilde{f}^{*}_{n},\cdots, \widetilde{f}^{K*}_{n}]$, $\boldsymbol{\vartheta}_{n}=[\vartheta^1_{n}, \cdots, \vartheta^K_{n}]$, and $\vartheta^k_{n}$ is defined as
\begin{equation}\label{vertana}
    \vartheta^k_{n} =\frac{\frac{\partial u^k_n(\textbf{a}_n, \widetilde{\textbf{f}}_n)}{\partial \widetilde{f}_n^k}}{\sqrt{\sum_{k=1}^{K} (\frac{\partial u^k_n(\textbf{a}_n, \widetilde{\textbf{f}}_n)}{\partial \widetilde{f}_n^k})^2}}.
\end{equation}
The robust game is $\widetilde{\mathcal{G}}=\{\mathcal{N},(\Psi)_{n\in \mathcal{N}},\widehat{\mathcal{A}}\}$.
\begin{proof}
See Appendix A.
\end{proof}

\textbf{Theorem 1}: For any set of system parameters and strategy space of users, there always exists an RNE for $\widetilde{\mathcal{G}}$.
\begin{proof}
From part 2) in Lemma 1, RNE is an instance of the generalized Nash equilibrium (GNE) (see (2) in \cite{QVIGNE}), and $\widetilde{\textbf{a}}^*$ is the RNE iff it is a solution to $QVI(\widehat{\mathcal{A}},\widetilde{\mathcal{F}})$, where $\widetilde{\mathcal{F}}(\textbf{a})=(\widetilde{\mathcal{F}}_n(\textbf{a}))_{n=1}^{N}$ and $\widetilde{\mathcal{F}}_n(\textbf{a})=-\frac{\partial \Psi_n(\textbf{a}_n, \textbf{a}_{-n})}{\partial \textbf{a}_n}$. Since $\widehat{\mathcal{A}}$ is a convex set and $\Psi_n(\textbf{a}_n, \textbf{a}_{-n})$ is a concave and continuous differentiable function with respect to $\textbf{a}_n$, the necessary convexity assumptions for the existence of a solution to $QVI$ hold (Theorem 1 in \cite{QVIGNE}), meaning that a RNE always exists.
\end{proof}

\subsection{RNE's Uniqueness Condition}
Since the closed form solution to (\ref{optrobsut}) cannot be obtained, the fixed-point algorithm and the contraction mapping cannot be applied as in \cite{Nash1,mihaelastructure} to derive the conditions for RNE's uniqueness. To overcome these difficulties, we show that the RNE can be considered as a perturbed NE of the NACG, and that the condition for RNE's uniqueness can be derived without a closed form solution to (\ref{optrobsut}).

\textbf{Lemma 2.} $\widetilde{ \mathcal{F}}(\textbf{a})$ is a perturbed bounded version of mapping $\mathcal{F}(\textbf{a})$.
\begin{proof}
See Appendix B.
\end{proof}

From Lemma 1, $\widehat{\mathcal{A}}$ is a closed and convex set, and form Lemma 2, $\widetilde{ \mathcal{F}}(\textbf{a})$ is a perturbed bounded mapping $\mathcal{F}(\textbf{a})$. Therefore, the RNE is a perturbed solution to $VI(\widehat{\mathcal{A}},\mathcal{F})$. Consequently, RNE's uniqueness condition can be obtained from the perturbed NE's uniqueness condition.

\textbf{Theorem 2.} When $\boldsymbol{\Upsilon}$ is a $P$ matrix, for any bounded value of
$\boldsymbol{\Delta}=[\varepsilon_1,\cdots,\varepsilon_N]$: 1) $\widetilde{\mathcal{G}}$ has a unique RNE; 2)The social utility at the RNE is always less than that at the NE, i.e., $\widetilde{u}^*\leq v^*$; 3) The distance between the strategy spaces at the RNE and at the NE is
    \begin{equation} \label{upperbound of variations}
    \|\textbf{a}^*-\widetilde{\textbf{a}}^*\|_2\leq \frac{\|\boldsymbol{\Delta}\|_2}{c_\text{sm}(\mathcal{F})}
\end{equation}
where $c_\text{sm}$ is the strong monotonicity constant for mapping $\mathcal{F}$.
\begin{proof}
See Appendix C.
\end{proof}
From Theorem 2, RACG's performance can be examined and compared to that of the ACG through the difference between the upper bound of users' strategies in (\ref{upperbound of variations}) at their equilibriums. Also, from (\ref{upperbound of variations}), the social utility's reduction at the RNE (compared to NE) can be approximated. Consider $\textbf{W}(\textbf{a})=(\textbf{W}^{k}(\textbf{a}))_{k=1}^{K}$, where $\textbf{W}^k$ is a $N \times N$ matrix whose elements are
\begin{eqnarray} \label{S}
 W^k_{nm} \equiv \left(\begin{array}{l}
 \frac{\partial v_n^k(a_{n}^{k}, f_{n}^{k})}{\partial a_{n}^k} \quad \quad \qquad\!\!\text{if}  \quad m=n  \\
\frac{\partial v_n^k(a_{n}^{k}, f_{n}^{k})}{\partial a_{m}^k} x_{nm}^k \quad \quad \text{if} \quad  m\neq n
\end{array} \right), \qquad m , n \in \mathcal{N}
\end{eqnarray}
In part 4 of Appendix C, we show that the difference between social utilities at the RNE and the NE is
\begin{equation}\label{diffrencebetweenustilitues}
    \| v^*-\widetilde{u}^*\|_2 \approx \| \textbf{W}(\textbf{a}^*)\| _2 \times \frac{\|\boldsymbol{\Delta}\|_2}{c_\text{sm}(\mathcal{F})}.
\end{equation}
When $\boldsymbol{\Upsilon}$ is a $P$ matrix, $\textbf{a}_{n}^{*}$ is the attractor for $VI(\widehat{\mathcal{A}}, \widetilde{\mathcal{F}})$ (Proposition 2.4.10 in \cite{PangVI}), i.e.,
\begin{equation}\label{upperboundofsolution}
   \lim_{\boldsymbol{\Delta}\rightarrow 0} \|\textbf{a}^{*}-\widetilde{\textbf{a}}^{*}\|_2 =0 \,\quad \quad \forall \textbf{a}^{*} \in \mathcal{A}, \quad \widetilde{\textbf{a}}^{*} \in \widehat{\mathcal{A}}, \nonumber
\end{equation}
meaning that when uncertainty approaches zero, the RNE converges to the NE.

\subsection{Numerical Validation}

Consider the power allocation problem (Example 1 in Table I), where the transmit power of user $n$ on sub-channel $k$ is $a_{n}^{k}$ in (\ref{An}). We have $-\nabla^{2}_{\textbf{a}_{n}} v_{n}(\textbf{a}_n, \textbf{f}_n)= \text{diag}(\frac{1}{(\bar{\sigma}_{n}^{k}+\sum_{m \in \mathcal{N}}a_{m}^{k}\bar{h}_{nm}^{k})^2})_{k=1}^{K}$. In this case, $\alpha_{n}^{\text{min}}=\min
(\frac{1}{\bar{\sigma}_{n}^{k}+\sum_{m \in \mathcal{N}}a_{m,k}^{\text{max}}\bar{h}_{nm}^{k}})^2 $, and $ - \nabla_{\textbf{a}_{m}\textbf{a}_{n}} v_{n}(\textbf{a}_n, \textbf{f}_n)= \text{diag}(\frac{\bar{h}_{nm}^{k}}{(\bar{\sigma}_{n}^{k}+\sum_{m \in \mathcal{N}}a_{m}^{k}\bar{h}_{nm}^{k})^2})_{k=1}^{K}$. Therefore, $\beta_{nm}^{\text{max}}= \max_{k \in \mathcal{K}}\frac{\bar{h}_{nm}^{k}}{(\bar{\sigma}_{n}^{k}+\sum_{m \in \mathcal{N}}a_{m,k}^{\text{min}}\bar{h}_{nn}^{k})^2}.$
For simplicity, assume that the normalized noise power in all sub-channels for each user are the same, and $a_{n,k}^{\text{min}}=0$. In such cases, the matrix $\boldsymbol{\Upsilon}$ is a $P$ matrix iff
\begin{equation}\label{uniqnessconditionsof gamepower control}
\min_{k \in \mathcal{K}}\frac{1}{(\bar{\sigma}_{n}^{k}+\sum_{m \in \mathcal{N}}a_{m,k}^{\text{max}}\bar{h}_{nm}^{k})^2}> \sum_{m \neq
    n}\max_{k \in \mathcal{K}}
    \frac{\bar{h}_{nm}^{k}}{(\bar{\sigma}_{n}^{k})^2}
    \quad\quad \forall n \in \mathcal{N},
\end{equation}
which is the uniqueness condition for the power control game. Otherwise, the power
allocation problem may have multiple Nash equilibria. As a numerical example for Theorem 1, we consider two users and two sub-channels, and assume $a_{n}^{\texttt{max}}=a_{n,k}^{\texttt{max}}=1$. To ensure that the monotonicity condition
(\ref{uniqnessconditionsof gamepower control}) of the power allocation game holds and $\boldsymbol{\Upsilon}$ is a $P$ matrix, the interference channel gain between users is chosen much less than the direct channel gain between transmitters and their receivers, i.e., $\bar{h}_{nm}^{k}<0.01$. In our simulation, the matrix $\boldsymbol{\Upsilon}$ is \[
\boldsymbol{\Upsilon} = \left[ {\begin{array}{cccc}
  1.5432 & -0.016\\
  -0.0012 & 1.221
 \end{array} } \right]
\]
Since (\ref{uniqnessconditionsof gamepower control}) is satisfied for the above matrix, it is a strictly monotone matrix. Figs. \ref{1} (a) and \ref{1} (b) show the mappings $\mathcal{F}_1(\textbf{a})$ and
$\mathcal{F}_2(\textbf{a})$ for the allocated power levels in subchannel 1 for users 1 and 2, respectively.

\begin{figure}
\centering
\includegraphics [height=5cm,width=8.5cm] {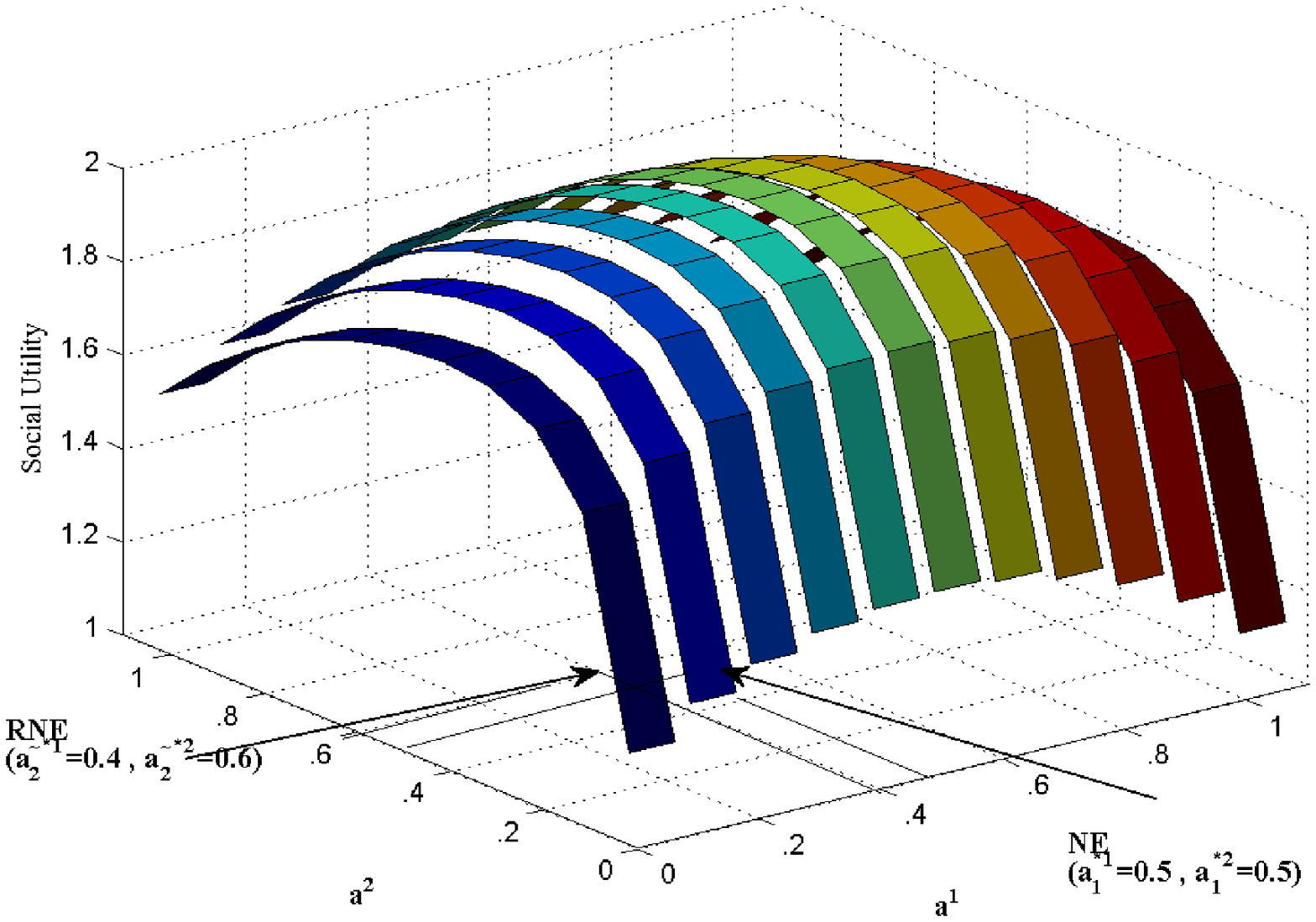}
\caption{The social utility for the power control game when $\boldsymbol{\Upsilon}$ is a $P$ matrix.}{\label{2}}
\end{figure}

For $n \in \{1,2\}$, the mapping $\mathcal{F}_n(\textbf{a})$ is obtained by considering a fixed transmit power of the other user, and as we show in Figs. \ref{1} (a) and \ref{1} (b), it is monotone. Fig. \ref{2} shows the surface of the social utility of users 1 and 2 for $\textbf{a}^{1}$ and $\textbf{a}^{2}$ as their strategies in sub-channel 1 and 2, respectively. We consider $\varepsilon_n\leq 0.8$ for both users. In this case, our simulations show that: 1) Both the NE and the RNE are unique; 2) The social utility at NE is $v^*_1+v^*_2=1.6$. The social utility at the RNE is $\widetilde{u}^*_1+\widetilde{u}^*_2=0.585$, which is less than the social utility at the NE, as expected from part 1 in Theorem 1; and 3) At the NE, the allocated power for users 1 and 2 are $(a_1^{*1}=0.5,a_1^{*2}=0.5)$ and $(a_2^{*1}=0.4,a_2^{*2}=0.6)$, respectively; and the RNEs for users 1 and 2 are  $(\widetilde{a}_1^{*1}=0.4,\widetilde{a}_1^{*2}=0.6)$ and $(\widetilde{a}_2^{*1}=0.9,\widetilde{a}_2^{*2}=0.1)$. In this example, the upper bound in (\ref{upperbound of variations}) is $1.3115$, and the exact value of $\|\textbf{a}^*-\widetilde{\textbf{a}}^*\|_2$ is $0.7211$, which is less than its upper bound (\ref{upperbound of variations}). Consider $\|\textbf{W}^{k}(\textbf{a})\|_2=\sqrt{(\lambda_{\max}(\textbf{W}^{*k}(\textbf{a})\textbf{W}^{k}(\textbf{a}))}$, where $\textbf{W}^{*k}(\textbf{a})$ is the conjugate transpose of $\textbf{W}^{k}(\textbf{a})$ and $\lambda_{\max}$ is the maximum eigenvalue of the matrix. In our simulation $\lambda_{\max}(\textbf{W}^{*1}(\textbf{a})\textbf{W}^{1}(\textbf{a}))=0.9091$ and $\lambda_{\max}(\textbf{W}^{*2}(\textbf{a})\textbf{W}^{2}(\textbf{a}))=0.59$. From (\ref{diffrencebetweenustilitues}), the distance between the social utilities at the RNE and at the NE is $1.02$, and in simulation, this difference is to $1.015$. Note that all of the above results validate Theorem 2.

\section{Logarithmic Utility Function}

We now consider a special form of uncertainty region and utility function that are pertinent to Example 1 in Table I, to show that in such cases, \textit{VI} becomes very simple, and the strict monotonicity requirement is relaxed to the positive semidefiniteness of an affine mapping. Let the utility of each user be \cite{walrand}
\begin{equation}\label{utilitylogconvex}
 v_{n}^{k}(a_{n}^{k},f_{n}^{k} )= \left\{\begin{array}{c}
   \!\!\!\!\!\!\!\!\!\!\!\!\!\!\!\!\!\!\!\!\!\!\!\!\!\!\!\!\!\!\!\!\!\!\!\!\!\!\!\!\!\!\!\!\!\!\!\!\!\!\!\!\!\!\!\!\!\!\log(c_{n}^{k}+\frac{a_{n}^{k}}{f_{n}^{k}}), \,\qquad\text{if} \qquad\, \theta=1 \\
    \frac{(c_n^k+\frac{a_{n}^{k}}{f_{n}^{k}})^{\theta+1}}{\theta+1}, \,\qquad\qquad\text{if} \,\,\, -1<\theta<0 \: \qquad\text{or} \qquad \,\theta<-1  \\
  \end{array} \right.
\end{equation}
where $c_{n}^{k}$ is the fixed system parameter related to each dimension of each user. Assume that for the uncertain parameter of each user, we have $\widetilde{x}_{nm}^{k}=x_{nm}^{k}+\widehat{x}_{nm}^{k}$ where $\widetilde{x}_{nm}^{k}$, $x_{nm}^{k}$ and $\widehat{x}_{nm}^{k}$ are the actual value, the nominal value, and the error, respectively. The uncertainty region for each user is \begin{equation}\label{Linearf2}
\Re_{n}^{k}=
\{ \sqrt{
\sum_{m=1, m\neq n}^{N} (\frac{\widehat{x}_{nm}^{k}}{x_{nn}^{k}})^2} \leq
\epsilon_{n}^{k}  \}, \quad \quad \forall k
\end{equation}
where $\epsilon_{n}^{k}$ is the bound on the uncertainty region for user $n$ in sub-channel $k$. The above formulation models the power control game when uncertainty emanates from variations in the channel gain between transmitter $m$ and receiver $n$.

\textbf{Proposition 1.} For an ACG with utility function (\ref{utilitylogconvex}): 1) The NE is the solution to an affine \textit{VI} denoted by $AVI(\mathcal{A},\mathcal{M})$ where $\mathcal{M}(\textbf{a})=(\mathcal{M}_n(\textbf{a}))_{n=1}^{N}$, and
     \begin{equation}\label{propostion1AVI}
     \mathcal{M}_n(\textbf{a})=\textbf{w}_{n}+\sum_{m=1}^{N}\textbf{M}_{nm}\textbf{a}_m^{{\scriptsize{\textnormal{T}}}}
     \end{equation}
     where $ \textbf{w}_{n}=(\frac{y_{n}^{k}+c_n^k}{x_{nn}^k})_{k=1}^{K}$ and $\textbf{M}_{mn}=\mathrm{diag}(\frac{x^{k}_{nm}}{x^{k}_{nn}})_{k=1}^{K}$. 2) Consider a $N \times N$ matrix $\textbf{M}^{\mathrm{max}}$ whose elements are
\begin{eqnarray}\label{mmax}
    M^{\mathrm{max}}_{nm}= \left\{\begin{array}{c}
 \max_{k \in \mathcal{K}}  \frac{x^{k}_{nm}}{x^{k}_{nn}} \,\qquad\text{if} \qquad\, m\neq n\nonumber \\
\!\!\!\!0 \, \,\,\,\quad\quad\qquad\qquad\mathrm{otherwise}  \\
\end{array} \right.
 \end{eqnarray}
The NE is unique when
\begin{equation}\label{conditionproposition1}
\max_{n \in \mathcal{N}} \quad \|\textbf{a}_n\|_2> \sum_{m\neq n} M^{\mathrm{max}}_{nm} \|\textbf{a}_m\|_2, \quad  \forall \textbf{a}_n \in \mathcal{A}_n, \quad \forall n , m \in \mathcal{N}.
\end{equation}
\begin{proof}
See Appendix D.
\end{proof}
From Proposition 1, when the impact of users on each other is sufficiently low, the NE is unique. For the example of power control game, Proposition 1 implies that when interference between users is sufficiently low, the NE of NACG is unique \cite{Luoiterativewaterfilling,VIintroduction}. We now derive the RNE' uniqueness condition for such cases.

\textbf{Theorem 3.} For the utility function (\ref{utilitylogconvex}): 1) The \textit{AVI} mapping of the RNE is $\widetilde{\mathcal{M}}(\textbf{a})=(\widetilde{\mathcal{M}}(\textbf{a}))_{n=1}^{N}$, where \begin{equation}\label{AVIlog}
    \widetilde{\mathcal{M}}_n(\textbf{a})=\mathcal{M}_n(\textbf{a})+\widehat{\mathcal{M}}_n(\textbf{a}), \quad \forall n \in \mathcal{N},
 \end{equation}
 where $\widehat{\mathcal{M}}_n(\textbf{a}_{-n})=(\epsilon_{n}^{k}\|\textbf{a}_{-n}^k\|)_{k=1}^{K}$ and $\textbf{a}_{-n}^{k}=[a_{1}^k,\cdots,a_{n-1}^k,a_{n+1}^k,\cdots,a_N^k]$; 2) When (\ref{conditionproposition1}) holds, the RNE of $\widetilde{\mathcal{G}}$ is unique for any bounded $\epsilon_n^k$; 3) When (\ref{conditionproposition1}) holds, the total utility of each user at the RNE is always less than that at the NE, and the upper bound on the strategy space of each user is
$\|\textbf{a}^*-\tilde{\textbf{a}}^*\|\leq\frac{\|\boldsymbol{E}\|_2^{2}}{\lambda_\text{min}(\textbf{M}^{\text{max}})}$,
where
\begin{eqnarray} E_{nm}= \left\{\begin{array}{c}
 \|\boldsymbol{\epsilon}_{n} \|_\infty\quad \text{if} \,\,\, m = n\nonumber \\
\:\,\:0 \, \,\quad \quad\quad\mathrm{otherwise}  \\
\end{array} \right.
\end{eqnarray}
where $\lambda_\text{min}$ is the minimum eigenvalue of matrix $\textbf{M}^{\text{max}}$, $\boldsymbol{\epsilon}_{n}=[\epsilon_{n}^{1}, \cdots, \epsilon_{n}^{K}]$ and $\|.\|_\infty$ is the maximum element of the vector.
\begin{proof}
See Appendix E.
\end{proof}
\textbf{Remark 3.} From Theorems 2 and 3, RNE's uniqueness condition is not related to the size of the uncertainty region. Theorems 2 and 3 show that for a closed, bounded, and convex uncertainty region, RNE's uniqueness condition is the same as NE's uniqueness condition. By rearranging \textit{AVI} of the RNE for the utility function (\ref{utilitylogconvex}), the best response solution of the RACG is
\begin{equation}\label{bestresponserobust2}
\widetilde{a}_{n}^{k}=[\lambda_n^{\frac{1}{\theta}}-\frac{c_{n}^{k}+y_{n}^{k}}{x_{nn}^{k}}-\sum_{m\neq
n} \frac{x_{nm}^{k}a_{m}^{k}}{x_{nn}^k}- \epsilon_n^k \|\textbf{a}_{-n}^k \|]_{a_{nk}^{\text{min}}}^{a_{nk}^{\text{max}}},
\end{equation}
which is similar to (13) in \cite{Robustnew} for the power control game in spectrum sharing environments. From Theorem 2 in \cite{Robustnew}, the RNE's uniqueness condition is related to the size of the uncertainty region. But if we use \textit{VI} to analyze the RNE, its uniqueness condition is independent of the size of the uncertainty region.

\section{Distributed Algorithms}

To develop a distributed algorithm for this class of games, we use the proximal-response map $\widehat{\textbf{a}}$. Consider $\widehat{\textbf{a}}=[\widehat{\textbf{a}}_1,\cdots,\widehat{\textbf{a}}_N]$, where $\widehat{\textbf{a}}$ is the solution to following optimization problem with respect to $\textbf{b}$
\begin{equation}\label{proximalresponsemap}
   \widehat{\textbf{a}}(\textbf{b})\triangleq \max_{\textbf{a} \in \widehat{\mathcal{A}}} [\sum_{n=1}^{N} \Psi_n(\textbf{a}_n,\textbf{b}_{-n})-\frac{1}{2} \|\textbf{a}-\textbf{b}\|_2^2],
\end{equation}
where $\textbf{b}=[\textbf{b}_1,\cdots,\textbf{b}_N]$ belongs to $\widehat{\mathcal{A}}$. The above problem can be decomposed into $N$ subproblems as
\begin{equation}\label{proximalresponsemapn}
 \widehat{\textbf{a}}_n(\textbf{b})= \max_{\textbf{a}_n \in \mathcal{A}_n(\textbf{a}_{-n})} [\Psi_n(\textbf{a}_n,\textbf{b}_{-n})-\frac{1}{2} \|\textbf{a}_n-\textbf{b}_n\|_2^2],  \quad , \textbf{b}\in \widehat{\mathcal{A}}(\textbf{a}_{-n}) , \quad \forall n \in \mathcal{N},
\end{equation}
In this way, a distributed iterative algorithm is developed. Consider $\widehat{\textbf{a}}_n$ and $\textbf{b}_n$ as solutions for user $n$ in the current and pervious iterations, respectively. When a user is informed about the other users' solutions before updating its strategy, it can obtain the solution to (\ref{proximalresponsemapn}). The distributed algorithm based on proximal response map is summarized in Table \ref{distributedalgorith}, where users update their transmit strategies at discrete instances $t$ denoted by $\mathcal{T}=[1,\cdots,T]$, $\textbf{a}_n(t)$ denotes the transmission strategy of user $n$ at iteration $t$, and $\textbf{f}_n(t)$ is the observation of user $n$ at $t$ measured by its receiver and sent to the transmitter.

In Theorem 4 below, we obtain the conditions for convergence of the iterative algorithm. 

\textbf{Theorem 4.} As $T \rightarrow \infty$, the distributed algorithm in Table \ref{distributedalgorith} converges to the unique RNE from any initial strategy $\textbf{a}_n(0)$, if $\boldsymbol{\Upsilon}$ is a $P$ matrix, and $ \frac{\partial^3 v_{n}^{k}(\textbf{a}_{n}, f_{n}^{k})}{\partial^2 a^k_n \partial f^k_n}=\frac{\partial^3 v_{n}^{k}(a_{n}, f_{n}^{k})}{\partial a^k_n \partial^2 f^k_n}=0$.
\begin{proof}
See Appendix F.
\end{proof}
Note that the condition $\frac{\partial^3 v_{n}^{k}(\textbf{a}_{n}, f_{n}^{k})}{\partial^2 a^k_n \partial f^k_n}=0$ holds for the Jackson network. For the power control game, when $\boldsymbol{\Upsilon}$ is a $P$ matrix, interference in the system is very low, and consequently, the signal to interference and noise ratio of each usr is high. For this case, $\frac{\bar{h}_{nn}^{k}a_{n}^{k}}{f_{n}^{k}}>>1$, and the utility function of each user is  $v_n(\textbf{a}_n,\textbf{f}_n)=\sum_{k=1}^{K} \log(\frac{\bar{h}_{nn}^{k}a_{n}^{k}}{f_{n}^{k}})$ which also meets $\frac{\partial^3 v_{n}^{k}(\textbf{a}_{n}, f_{n}^{k})}{\partial^2 a^k_n \partial f^k_n}=0$. As we see in Theorem 4, the distributed algorithm converges to the unique NE when $\boldsymbol{\Upsilon}$ is a $P$-matrix irrespective of the size of the uncertainty region, so long as the uncertainty region is closed and convex.

\textbf{Remark 4.} In Lemma 1, we showed that $\Psi_n=u_n(\textbf{a}_n, \textbf{f}_n- \varepsilon_n \boldsymbol{\vartheta}_{n})$ is concave. Also the proximal response map is strictly convex. Therefore, the Lagrange function can be used to derive the solution to (\ref{proximalresponsemapn}) for each user at each iteration as
\begin{equation}\label{Lagrangeofproximal}
    L_{n}(\textbf{a}_n, \mu_n)=u_n(\textbf{a}_n, \textbf{f}_n- \varepsilon_n \boldsymbol{\vartheta}_{n})-\frac{1}{2} \|\textbf{a}_n(t)-\textbf{a}_n(t-1)\|_2^2 - \mu_{n}(\sum_{k=1}^{K}a_n^k-a_n^{\text{max}}),
\end{equation}
where $\mu_n$ is the lagrange multiplier for user $n$ that satisfies
\begin{equation}\label{mu}
    \mu_{n}(\sum_{k=1}^{K}a_n^k-a_n^{\text{max}})=0,
\end{equation}
The solution to (\ref{Lagrangeofproximal}) with respect to $a^k_n$ is
\begin{eqnarray}\label{solution}
   && \frac{ \partial L_{n}(\textbf{a}_n, \mu_n)}{\partial a_{n}^{k}}= \\ &&\frac{\partial u_n^k (a_n^k, f_n- \varepsilon_n \vartheta^k_{n})}{\partial a_n^k} -\varepsilon_n \frac{\partial u_n^k (a_n^k, f_n- \varepsilon_n \vartheta^k_{n})}{\partial f_n^k} \frac{\vartheta^k_{n}}{\partial a_n^k}-\mu_n-(a_n^k(t)-a_{n}^k(t-1))=0 ,  \quad \forall k \in \mathcal{K}, \nonumber
\end{eqnarray}
User $n$ solves (\ref{solution}) to obtain $\textbf{a}_n(t)$ in each iteration. For (\ref{utilitylogconvex}), the proximal map's solution is
\begin{equation}\label{propostion1AVI}
     \mathcal{M}_n(\textbf{a}(t))=\textbf{w}_{n}+\sum_{m=1}^{N}\textbf{M}_{nm}\textbf{a}^{{\scriptsize{\textnormal{T}}}}_{m}(t-1)+\mathcal{I}_n,
\end{equation}
where $\mathcal{I}_n=(a_{n}^k(t)-a_{n}^k(t-1))_{k=1}^K$. For example, the proximal map's solution to the power control game is \begin{equation}\label{bestresponserobust2proximalmap}
a_{n}^{k}(t)_=\frac{1}{2}[\lambda_n^{\frac{1}{\theta}}-\frac{c_{n}^{k}+y_{n}^{k}}{x_{nn}^{k}}-\sum_{m\neq
n} \frac{x_{nm}^{k}a_{m}^{k}(t-1)}{x_{nn}^k}- \epsilon_n^k \|\textbf{a}_{-n}^k(t-1) \|+a_n^k(t-1)]_{a_{nk}^{\text{min}}}^{a_{nk}^{\text{max}}},
\end{equation}

\section{Effects of Robustness on Social Utility for the Case of Multipls NEs}

So far, based on NE's uniqueness condition, we obtained RNE's uniqueness condition. Now we study the characteristics of RNEs when NACG has multiple NEs. In general, doing so is not straightforward since the \textit{VI} mapping for NACG is non-monotone and non-smooth, which makes it difficult to study the characteristics of the perturbed NEs (RNEs)  \cite{nonsmoothbook,nonsmoothproblem,Enclopediaofoptimizationprobelm}.

\begin{figure}
\centering
\includegraphics [height=5cm,width=8.5cm] {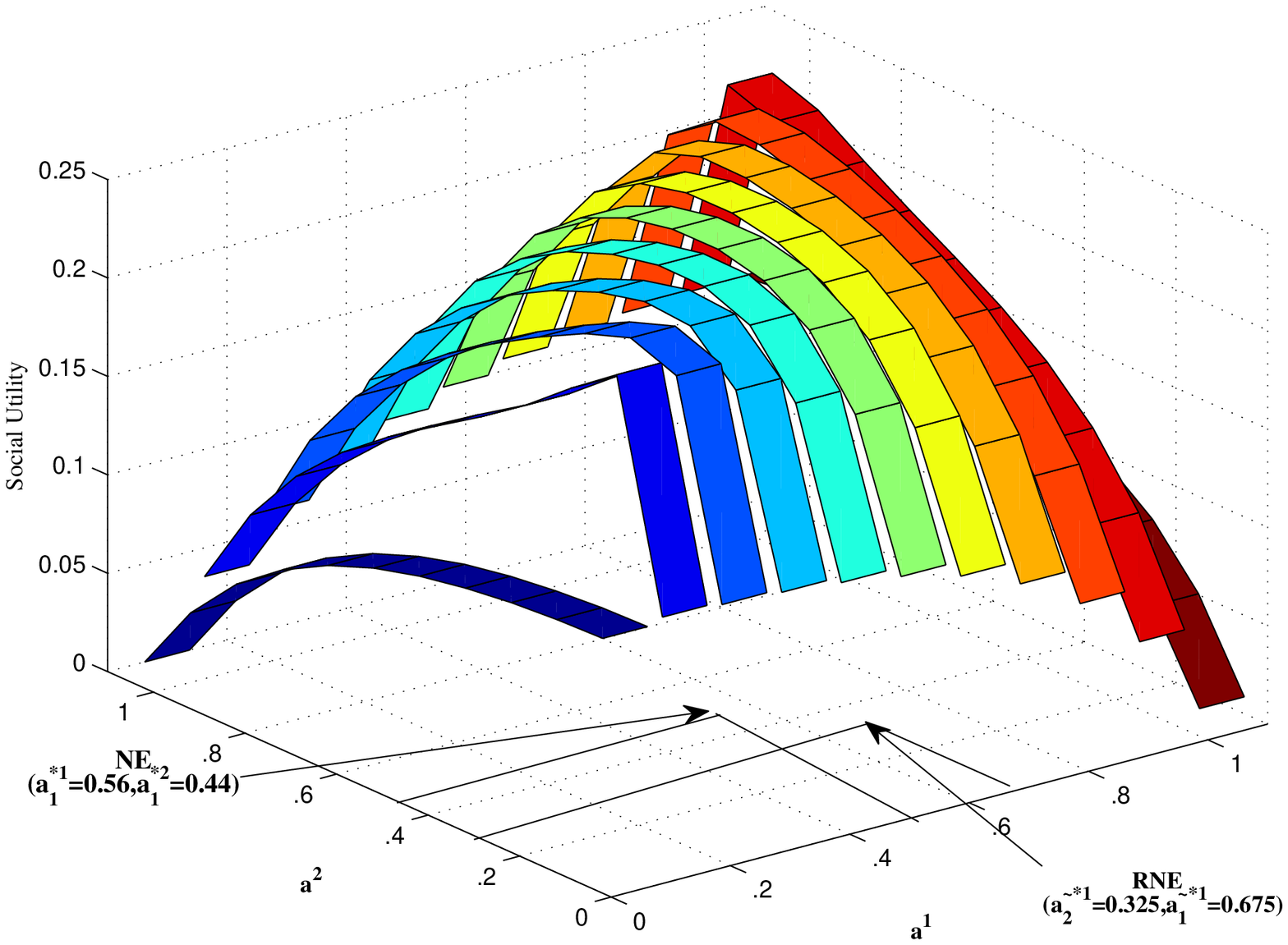}
\caption{The social utility for the power control game when $\boldsymbol{\Upsilon}$ is not a $P$ matrix.}{\label{4}}
\end{figure}

To compare the case of multiple NEs with that of a single NE, consider the power control game, when $\boldsymbol{\Upsilon}$ is not a $P$ matrix (e.g., $\bar{h}_{nm}^{k}>0.5$). The mapping $\mathcal{F}$ is non-monotone for both users. As we see in Fig. \ref{4}, there are multiple local optima on the surface of the utility function that correspond to multiple NEs for this game. In this case, at the nominal NE, the convergence points for users 1 and 2 are $(a_1^1=0.534,a_1^2=0.463)$ and $(a_2^1=0.417,a_2^2=0.583)$, respectively, and $v^*_1+v^*_2=3.0176$. When uncertainty is $\varepsilon_n<0.8$, the RNE converges to $(\widetilde{a}_1^1=0.556,\widetilde{a}_1^2=0.444)$ and $(\widetilde{a}_2^1=0.325,\widetilde{a}_2^2=0.675)$, and $\widetilde{u}^*_1+\widetilde{u}^*_2=3.077$. This example points out that introducing uncertainty may increase the social utility at the RNE when the NACG has multiple NEs, which is in line with simulation results in \cite{Robustnew,ProbabilisticIWFA,SaeedehIET}. This is because RNE relates to one local optima on the surface of the utility function, which is different from the local optima at the NE. Also, considering uncertainty results in users interfering less with each other at the RNE compared to the NE  \cite{SaeedehIET}.

This observation shows the benefit of implementing RACG in communication systems which may increase the social utility as compared to that of NACG. But, obtaining the conditions under which the social utility increases is not easy. This is because utility function of each user is a non-linear function with respect to its uncertainty region and the other users' uncertainty regions. So, we focus on a special case where the strategy of each user is a decreasing function of the bound of uncertainty region. In Proposition 2 below, we obtain the condition for increasing the social utility of the RACG as compared to that of the NACG.

\textbf{Proposition 2.} When $\textbf{W}$ is a semi negative definite matrix and $\nabla_{\varepsilon_n} \textbf{a}_{n}<0$ for all users, the social utility at the RNE is higher than that at the NE.
\begin{proof}
See Appendix G.
\end{proof}
Proposition 2 implies that when the reduction in the social utility due to the decrease in user's strategies is less than their increase in the social utility due to the decrease in other usrs' strategies, introducing robustness in the game increases the social utility. Note that this is one case in which the social utility at RNE is higher than the social utility at the corresponding NE, and there may be other cases as well.

\textbf{Remark 5.} When the solution of affine \textit{VI} is a monotone decreasing function of $\varepsilon_n$, the \textit{AVI} mapping is a semi-negative matrix (See Appendix H). For the power control game, Proposition 2 is simplified to
\begin{equation}\label{conditionproposition1forpowercontrolgame}
\max_{n \in \mathcal{N}} \quad \|\textbf{a}_n\|_2 < \sum_{m\neq n} M^{\mathrm{max}}_{nm} \|\textbf{a}_m\|_2, \quad  \forall \textbf{a}_n \in \mathcal{A}_n, \quad \forall n , m \in \mathcal{N}.
\end{equation}

This means that when all interference channel gains are sufficiently greater than the direct channels gains, introducing robustness increases the social utility. This is an opportunistic phenomenon in robust games when the game encounters multiple NEs. In order to benefit from this and increase the social utility, we propose an opportunistic distributed algorithm in Table \ref{tableopportunesticalgorithm}. Obviously, checking the conditions of Proposition 2 in a distributive manner is not easy. In addition the social utility may increase under other conditions. Therefore, all users play the game $\mathcal{G}$ without considering the uncertainty. If $\alpha_{n}^{\text{min}}< \sum_{m\neq n} \beta_{m}^{\text{max}}$ for user $n$, i.e., the NACG has multiple NEs, users assume uncertainty in their observations. When their utilities increase, they expand their uncertainty regions. Otherwise, they interrupt the algorithm. In this way, all users make an effort to escape from their local optima in a distributed manner by playing the robust game. To implement this algorithm, we assume that users update their transmit strategy at discrete time slots $t_1=1,2,\cdots, T_1$ with duration of $\boldsymbol{\tau}$. The vectors $\textbf{a}_n(t_1)$ and $\textbf{f}_n(t_1)$ are the transmit strategy and the observation of each user at the end of the iteration time $t_1$. Besides, users exchange the values of $\textbf{a}_n(t_1)$ and $\widetilde{u}_n $ at the end of each iteration.

\section{Simulation Results}

We use simulations in the two examples in Table I to provide an insight into the performance of $\widetilde{\mathcal{G}}$ for different bounds on uncertainty region as compared to that of $\mathcal{G}$. In the following simulations, the value of $\varepsilon_n$ is normalized to the nominal value of $\textbf{f}_n$, i.e.,
$\varepsilon_n=\frac{\|\widetilde{\textbf{f}}_{n}-\textbf{f}_n\|}{\|\textbf{f}_n\|}$,
each uncertainty region is considered as a linear norm with order 2, i.e., an ellipsoidal region, and uncertainty for all users is assumed to be the same, denoted by $\varepsilon$.

\subsection{Power Control Game}

\begin{figure}
\centering
\includegraphics [height=5cm,width=8.5cm] {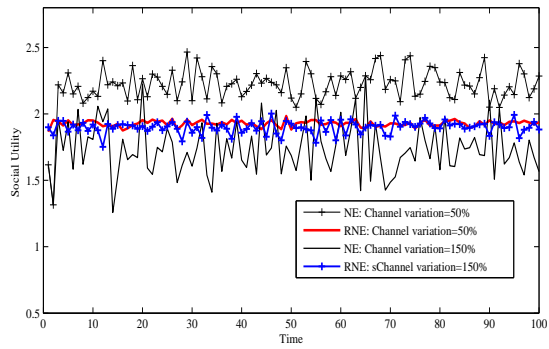}
\caption{The impact of channel variations in the robust and the non-robust games. }{\label{5}}
\end{figure}

For the power control game, we begin by studying the effect of uncertainty on its performance in both robust and non-robust approaches in terms of utility variations at their equilibria. To do so, we consider $N=3$ users and $K=16$, and the amount of uncertainty is assumed to be $\varepsilon=50\%$ at the RNE. After convergence to the RNE and to the NE, the system parameter varies uniformly from $50\%$ to $150\%$, which causes variations in the utility of each user at the NE and at the RNE. Variations in the social utility are shown in Fig. \ref{5}. Note that the social utility varies considerably at the NE of the nominal game for both values of $[50\%, 150\%]$, meaning that communication is very unreliable from the user's perspective. In contrast, the total system utility at the RNE of the robust game is stable for both cases. Note that although we assumed $\varepsilon=50\%$ in the RACG, reliable transmissions is provided even at values higher than $\varepsilon=50\%$, e.g., up to $150\%$.

\begin{figure}
\centering
\includegraphics [height=5cm,width=8.5cm] {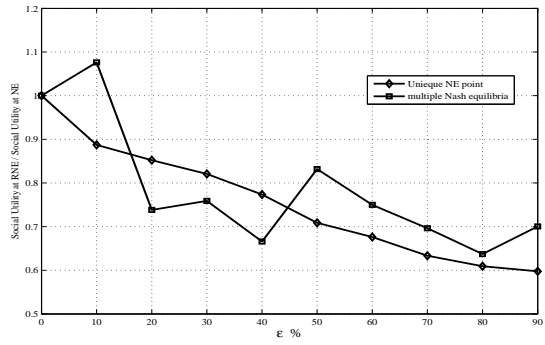}
\caption{RNE and NE for different uncertainty regions for unique NE and multiple NEs.}{\label{6}}
\end{figure}

\begin{figure}
\centering
\includegraphics [height=5cm,width=8.5cm] {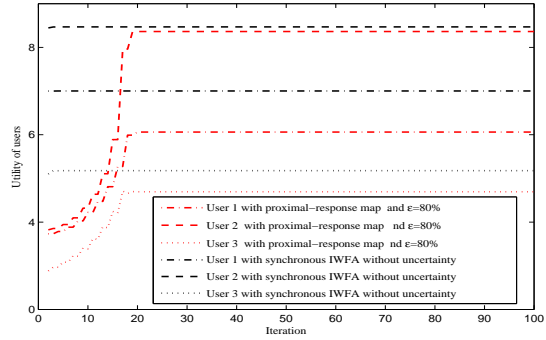}
\caption{Convergence times of the proximal response map and the IWFA.}{\label{7}}
\end{figure}

Next, in Fig. \ref{6}, we compare the effect of uncertainty when Theorem 2 holds with that of the case when it does not hold, in terms of the ratio of the social utility at the RNE and at the NE for different amounts of uncertainty. Simulations are performed for Rayleigh fading channels for bounded and uniformly generated errors for each cross sub-channel gain. To satisfy the NE's uniqueness condition, the channel gains are obtained in such a way that $\bar{h}_{nm}^{k}<0.01$, and for multiple NEs, we set $\bar{h}_{nm}^{k}>0.5$. The ratio of the social utility in Fig. \ref{6} is obtained by averaging over 100 channel realizations. Note that when the NE is unique, the total throughput of the system gradually decreases, but for the case of multiple NEs, no uniformity is observed. For example, when $\varepsilon= 10\%$, the total achieved utility is higher at the RNE as compared to that at the NE; and when $\varepsilon=20\%$, the ratio falls substantially. The trend is not monotonic for different values of uncertainty, e.g., the social utility at the RNE is higher for $\varepsilon= 50\%$ as compared to those for $\varepsilon= 40\%$ and $\varepsilon=60\%$. In Fig. \ref{7}, the throughput of the users versus iteration numbers are shown for the proximal response map when $\varepsilon=80\%$. Note that as shown in this figure, the convergence time of the proximal response map is longer than that of the IWFA without any  uncertainty for users.

\begin{figure}
\centering
\includegraphics [height=5cm,width=8.5cm] {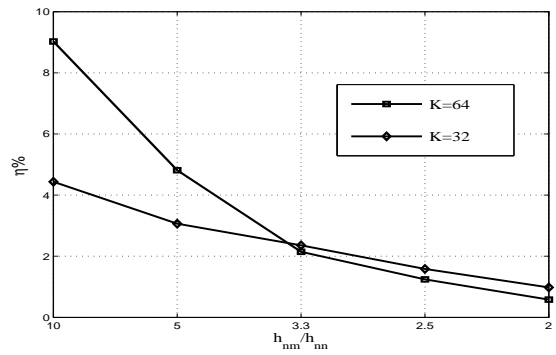}
\caption{Performance of the opportunistic algorithm in the power control game.}{\label{8}}
\end{figure}

Finally, in Fig. \ref{8}, we compare the performance of the opportunistic approach with that of the nominal game using
\begin{equation}\label{opportunesticapproach}
   \eta=\frac{\sum_{n \in \mathcal{N}} u^\text{OP}_n-v^*}{v^*},
\end{equation}
where $u^\text{OP}_n$ is the achieved utility of user $n$ by utilizing the opportunistic algorithm at the end of the algorithm. The value of $\eta$ is obtained by averaging over 1200 channel realizations for  $N=[4,6,8]$, $K=[32,64]$ for different values of $\bar{h}_{nm}^{k}$. Note that for very high interference levels, i.e., when $\bar{h}_{nm}^{k}\gg 1$, the efficiency of the proposed opportunistic approach is considerably higher as compared to that of moderate interference levels, i.e., when $\bar{h}_{nm}^{k}\geq 1$. Thus, the opportunistic algorithm performs significantly better in high interference levels when the number of sub-channels is high.

\subsection{Jackson Networks: M/M/1}

\begin{figure}
\centering
\includegraphics [height=5cm,width=11.5cm] {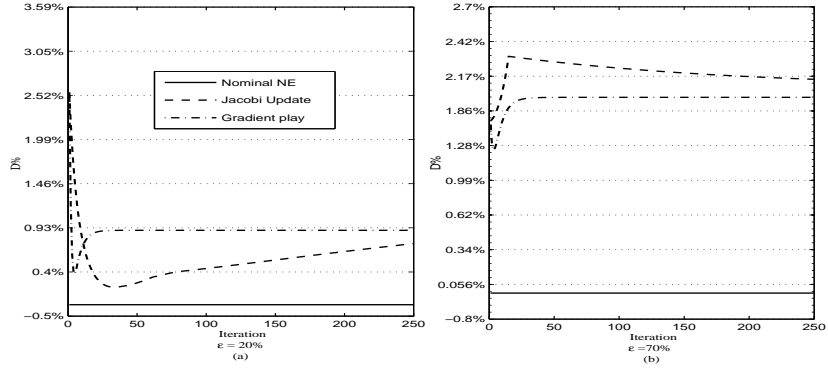}
\caption{The impact of uncertainty on the total delay of the Jackson network when (a)
$\boldsymbol{\varepsilon}=20\%$ and (b) $\boldsymbol{\varepsilon}=70\%$ for the gradient play and the Jacobi update \cite{mihaelastructure}.}{\label{9}}
\end{figure}

\begin{figure}
\centering
\includegraphics [height=5.5cm,width=10.5cm] {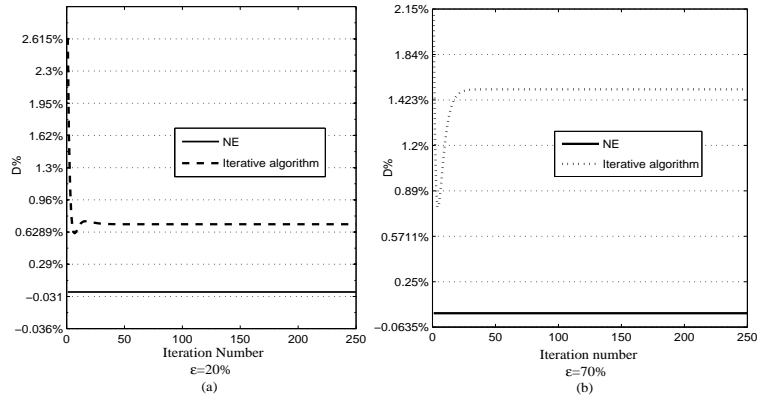}
\caption{Convergence to RNE for the proximal response map when (a)
$\boldsymbol{\varepsilon}=20\%$ and (b) $\boldsymbol{\varepsilon}=70\%$ .}{\label{10}}
\end{figure}

To show the effect of uncertainty on the system performance in Jackson networks, we consider a network with $N = 5$ nodes and $K = 3$ traffic classes. Figs. \ref{9}(a) and \ref{9}(b) show the effect of uncertainty in $\psi_{n}^{k}$ on the convergence of Jacobi scheme and the gradient approach for reaching the NE. In this figures, we show the percentage of increasing the total delay in the network under perturbation as compared to that at the NE, i.e., $D=\frac{\breve{d}-d^*}{d^*}\%$, where $\breve{d}$ is the total delay under perturbation and $d^*$ is the delay at the NE. Note that when robustness is not applied, neither of the two algorithms converge to the NE, and increasing uncertainty increases the queuing delay. Figs. \ref{10} (a) and \ref{10} (b) show the RNE for $\varepsilon=20\%$ and $\varepsilon=70\%$, respectively, by using the proximal response map. Note that in both cases, the RNEs converge to the vicinity of the nominal NE as compered to Figs. \ref{9}(a) and \ref{9} (b). For example, when $\varepsilon=70\%$, the total delay increases up to $2\%$ as compared to that at the NE, while the total delay is about $1.423\%$ at the RNE. Therefore, RNEs are stable, and the iterative algorithm converges very fast. Note that proximal response map, the convergence rate increases when uncertainty region is expanded.

\begin{figure}
\centering
\includegraphics [height=5cm,width=8.5cm] {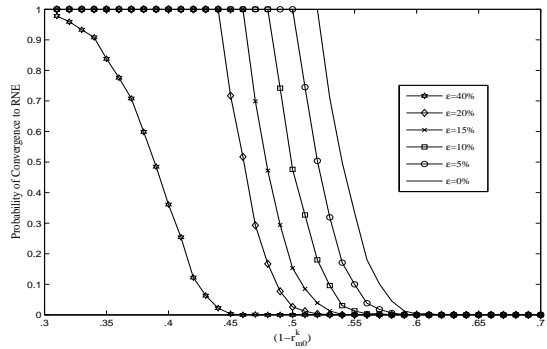}
\caption{Probability of convergence to RNE when $\boldsymbol{\Upsilon}$ is a $P$ matrix for different amounts of uncertainty versus $(1-r_{m0}^k)$ for $N=5$ and $K=3$. }{\label{11}}
\end{figure}

Fig. \ref{11} shows the probability of RNE's convergence,
versus the total routing probability $1-r^{k}_{m0}$ for different uncertainty regions. Note that by increasing uncertainty, the system converges to the RNE for a smaller value of $(1-r^{k}_{m0})$ as compared to that of the NE (i.e., $\varepsilon=0$). For example, if
$\varepsilon=40\%$, only for $(1-r^{k}_{m0})<0.3$, the system converges to its equilibrium, while for
$\varepsilon=10\%$, the value of $(1-r^{k}_{m0})$ can be up to $0.5$ for convergence to the RNE. Fig. \ref{11}
shows that the effect of uncertainty is more profound in a network with a high probability of packet loss, causing poor performance, i.e., large delays in the network. Therefore, from a practical perspective, when the system encounters a high degree of uncertainty, all nodes should reduce their probability of packet drops and increase their routing probability.

\section{Conclusions}

We studied the RNE for a wide range of problems in communication systems and networks when users' impact on each other can be modeled as an additive function of system parameters and users' strategies. In this game, since the user's observations of these impacts are uncertain, each user optimizes its utility using worst-case robust optimization. We showed that the theory of finite-dimension \textit{VI} and its sensitivity analysis can be used to obtain the conditions for the existence and uniqueness of the RNE when there is no closed-form solution to the optimization problem of each entity. We also proposed a distributed algorithm for reaching the RNE. In the case of multiple NEs, simulations showed that at a RNE, the social utility may be higher than that at the NE, and we derived the conditions for observing such a case. To benefit from this, we proposed a distributed algorithm to improve the social utility at the RNE. Simulations confirmed our analysis for two examples, namely, the power control in interference channels, and delay minimization in Jackson networks.

\begin{appendices}

\section{Proof of Lemma 1}
1) $\textbf{f}_{n}$ is a linear function of other users' strategies and system parameters. Besides, the norm function is a convex function bounded to $\varepsilon_n$. Therefore, $\widehat{\mathcal{A}}_n(\textbf{a}_{-n})$ is a convex, bounded and close set. 2) To prove the concavity of (\ref{psi2}) with respect to $\textbf{a}_n$, consider $\Psi_n(\textbf{a}_n^1)$, and $\Psi_n(\textbf{a}_n^2)$ for any positive value $\mu \in [0,1]$. We have
\begin{eqnarray}
 \Psi_n(\mu\textbf{a}_n^1+ (1-\mu)\textbf{a}_n^2)& =& \min_{\widetilde{\textbf{f}}_{n} \in \mathcal{R}_n(\textbf{a}_{-n})} u_n(\mu\textbf{a}_n^1+ (1-\mu)\textbf{a}_n^2,\widetilde{\textbf{f}}_{n}) \label{ps1}\\
& \geq&  \min_{\widetilde{\textbf{f}}_{n} \in \mathcal{R}_n(\textbf{a}_{-n})} \mu u_n(\textbf{a}_n^1,\widetilde{\textbf{f}}_n
 )+ (1-\mu) u_n(\textbf{a}_n^2, \widetilde{\textbf{f}}_{n}) \label{ps2}\\
 &\geq&  \min_{\widetilde{\textbf{f}}_{n} \in \mathcal{R}_n(\textbf{a}_{-n})} \mu u_n(\textbf{a}_n^1,\widetilde{\textbf{f}}_{n})+ (1-\mu)\min_{\widetilde{\textbf{f}}_{n} \in \mathcal{R}_n(\textbf{a}_{-n})} u_n(\textbf{a}_n^2, \widetilde{\textbf{f}}_{n})\\
&=&  \mu \Psi_n(\textbf{a}_n^1)+ (1-\mu)\Psi_n(\textbf{a}_n^2),
\end{eqnarray}
Note that inequality (\ref{ps2}) is based on concavity of $u_n(\textbf{a}_n,\widetilde{\textbf{f}}_n)$ with respect to $\textbf{a}_n$. Therefore, $\Psi_n(\textbf{a}_n^2)$ is a concave function of $\textbf{a}_n$. Based on concavity of $\Psi_n(\textbf{a}_n^2)$, the Lagrange dual function of (\ref{psi2}) for the uncertainty region is
\begin{equation}\label{lagrangedualfunction}
    L(\textbf{a}_n,\widetilde{\textbf{f}}_{n}, \lambda_n)= \sum_{k=1}^{K} u^{k}_n(a_n^k,\widetilde{f}^{k}_{n}) - \lambda_n (\varepsilon^n - \sum_{k=1}^{K} (\widetilde{f}_{n}^{k}-f_{n}^{k})^2)
 \end{equation}
where $\lambda_n$ is the nonnegative Lagrange multiplier that satisfies (\ref{III-1}), i.e.,
  \begin{equation}\label{lagrangemultiplier}
\lambda_n \times (\varepsilon_n^2 - \sum_{k=1}^{K} (\widetilde{f}_{n}^{k}-f_{n}^{k})^2)=0
 \end{equation}
The solution to (\ref{lagrangedualfunction}) for $\widetilde{f}_{n}^{k}$ can be obtained by the optimality condition of the optimization problem without the constraint \cite{boydconvexbook}, i.e., $\frac{\partial L(\textbf{a}_n,\boldsymbol{\widetilde{f}}_{n}, \lambda_n)}{\partial \widetilde{f}^{k}_n}=0$, which is equivalent to
\begin{equation}\label{solution2}
    \frac{\partial u^k_n(a^k_{n},\widetilde{f}^k_{n})}{\partial \widetilde{f}^k_{n}}= -2 \lambda_n \times (\widetilde{f}_n^k-f_n^k)  \qquad \forall k \in \mathcal{K}
\end{equation}
Considering (\ref{solution2}) in (\ref{lagrangemultiplier}), the uncertain parameter is
 $ \widetilde{\textbf{f}}^{*}_{n}=\textbf{f}_n- \varepsilon_n \boldsymbol{\vartheta}_{n}$,
where $\widetilde{\textbf{f}}_n^*=[\widetilde{f}^{*}_{n},\cdots, \widetilde{f}^{K*}_{n}]$, $\boldsymbol{\vartheta}_{n}=[\vartheta^1_{n}, \cdots, \vartheta^K_{n}]$, and $\vartheta^k_{n}$ is
\begin{equation}\label{vertana}
    \vartheta^k_{n} =\frac{\frac{\partial u^k_n(\textbf{a}_n, \widetilde{\textbf{f}}_n)}{\partial \widetilde{f}_n^k}}{\sqrt{\sum_{k=1}^{K} (\frac{\partial u^k_n(\textbf{a}_n, \widetilde{\textbf{f}}_n)}{\partial \widetilde{f}_n^k})^2}}
\end{equation}
Using $\vartheta^k_{n}$ in the utility function $u_n$, we have
\begin{equation}\label{Psi2}
     \Psi_n(\textbf{a}_n, \textbf{a}_{-n})=u_n(\textbf{a}_n, \textbf{f}_n)|_{ \widetilde{\textbf{f}}^{*}_{n}=\textbf{f}_n- \varepsilon_n \boldsymbol{\vartheta}_{n}},
\end{equation}
Comparing (\ref{Psi2}) with $v_n(\textbf{a}_n, \textbf{f}_n)$ indicates that the difference between $\Psi_n$ and the utility function of the nominal game is the extra term  $\varepsilon_n \boldsymbol{\vartheta}_{n}$. From \textbf{A2}, $\varepsilon_n \boldsymbol{\vartheta}_{n}$ is continuous. Therefore   $\Psi_n(\textbf{a}_n, \textbf{a}_{-n})$ is continuous with respect to $\textbf{a}_n$. The derivative of $\Psi_n$ with respect to $\textbf{a}_n$ is
\begin{eqnarray}\label{derivativepsi}
     \nabla_{\textbf{a}_n}\Psi_n(\textbf{a}_n, \textbf{a}_{-n})&=&  \nabla_{\textbf{a}_n} u_n(\textbf{a}_n, \textbf{f}_n- \varepsilon_n \boldsymbol{\vartheta}_{n}) + (\nabla_{\widetilde{\textbf{f}}_n} u_n(\textbf{a}_n, \textbf{f}_n- \varepsilon_n \boldsymbol{\vartheta}_{n} ) \times \textbf{1}_{K}) \times  \nabla_{\textbf{a}_n} \widetilde{\textbf{f}}_n \\&=& \nabla_{\textbf{a}_n} u_n(\textbf{a}_n, \textbf{f}_n- \varepsilon_n \boldsymbol{\vartheta}_{n}) - \varepsilon_n (\nabla_{\widetilde{\textbf{f}}_n}  u_n(\textbf{a}_n, \textbf{f}_n- \varepsilon_n \boldsymbol{\vartheta}_{n})\times \textbf{1}_K) \times \nabla_{\textbf{a}_n} \boldsymbol{\vartheta}_{n}, \label{derivativepsi3}
\end{eqnarray}
Where $\textbf{1}_K$ is a $1 \times K$ vector whose elements are equal to one. The last term in (\ref{derivativepsi3}) contains $\frac{\partial^2 u_n^k}{\partial a_n^k \partial f_n^k}$. From \textbf{A3}, the term $\frac{\partial^2 u_n^k}{\partial a_n^k \partial f_n^k}$ exists. Therefore, $\Psi_n(\textbf{a}_n, \textbf{a}_{-n})$ is differentiable with respect to $\textbf{a}_n$. Now, the optimization problem of each user can be rewritten as $  \widetilde{u}_{n} = \max_{\textbf{a}_{n} \in \mathcal{A}_n \times \mathcal{R}_n(\textbf{a}_{-n})} \Psi_n(\textbf{a}_n, \textbf{a}_{-n})$. Therefore, the game can be reformulated as $\{\mathcal{N}, \Psi_n, \widehat{\mathcal{A}}\}$.
\section{Proof of Lemma 2}
For the RACG, we have $VI(\widetilde{\mathcal{F}}, \widehat{\mathcal{A}})$, and $\widetilde{\mathcal{F}}(\textbf{a})=(\widetilde{\mathcal{F}}_{n}(\textbf{a}))_{n=1}^{N}$, where $\widetilde{\mathcal{F}}_n(\textbf{a})$ is obtained by (\ref{derivativepsi3}) for user $n$. Let $\hat{\textbf{z}}_n=\{\hat{\textbf{x}}_{n}, \hat{\textbf{a}}_m \}$ denote variations in the system's of parameters and other users strategies for user $n$ where $\hat{\textbf{x}}_{n}$ and $\hat{\textbf{a}}_m$ are variations of $\textbf{x}_{n}$ and $\textbf{a}_m$, respectively. When $\varepsilon_n=0$, we have $\hat{\textbf{z}}_n=0$ and vice versa. From \textbf{A1} and \textbf{A2}, the mapping $\widetilde{\mathcal{F}}_n$ is continuous and differentiable around the uncertain parameters. We use the Taylor series of the uncertain parameter and write
\begin{equation}\label{tailorseriesofmappingF}
    \widetilde{\mathcal{F}}_n(\textbf{a})=[\widetilde{\mathcal{F}}_n(\textbf{a})]_{\hat{\textbf{z}}_n=0}+ \varepsilon_{n}[\nabla_{\hat{\textbf{z}}_n} \widetilde{\mathcal{F}}_n(\textbf{a})]_{\hat{\textbf{z}}_n=0} +[\sum_{i=2}^{\infty} \frac{1}{i!}  (\varepsilon_{n})^i(\nabla^i_{\hat{\textbf{z}}_n} \widetilde{\mathcal{F}}_n)]_{\hat{\textbf{z}}_n=0}
\end{equation}
For $\varepsilon_n=\hat{\textbf{z}}_n=0$, (\ref{tailorseriesofmappingF}) is equivalent to
\begin{eqnarray} \label{tailorseriesmappingf}
&& \!\!\!\!\!\!\!\!\!\!\!\!\!\widetilde{\mathcal{F}}_n(\textbf{a})  = -[\nabla_{\textbf{a}_n} u_n(\textbf{a}_n, \textbf{f}_n- \varepsilon_n \boldsymbol{\vartheta}_{n}) - \varepsilon_n (\nabla_{\textbf{f}_n} u_n(\textbf{a}_n, \textbf{f}_n- \varepsilon_n \boldsymbol{\vartheta}_{n}) \times \textbf{1}_K) \nabla_{\textbf{z}_n} \boldsymbol{\vartheta}_{n} ]_{(\varepsilon_n=0)}  \\ \label{tailorseriesmappingf2}&& \!\!\!\!\!\!\!\!\!\!\!\!\!\!\! -\frac{\varepsilon_n}{2}[\nabla_{\textbf{a}_n \boldsymbol{f_n}}^2 u_n(\textbf{a}_n, \textbf{f}_n- \varepsilon_n \boldsymbol{\vartheta}_{n}) \nabla_{\textbf{z}_n} \boldsymbol{f_n} - \varepsilon_n \nabla_{\textbf{f}_n \textbf{f}_n}^2 u_n(\textbf{a}_n, \textbf{f}_n- \varepsilon_n \boldsymbol{\vartheta}_{n}) \nabla_{\textbf{z}_n} \boldsymbol{\vartheta}_{n} ]_{(\varepsilon_n=0)} \\&& \!\!\!\!\!\!\!\!\!\!\!\!\!\!\!\!\!\!\!\! \label{tailorseriesmappingf3} - \frac{\varepsilon^2_n}{3!}[\nabla^3_{\partial \textbf{a}_n \boldsymbol{f_n}^2} u_n(\textbf{a}_n, \textbf{f}_n- \varepsilon_n \boldsymbol{\vartheta}_{n}) (\nabla_{\textbf{z}_n} \boldsymbol{f_n}) ^2 \times \textbf{1}_K^{{\scriptsize{\textnormal{T}}}} + \nabla_{\textbf{a}_n \boldsymbol{f_n}}^2 u_n(\textbf{a}_n, \textbf{f}_n- \varepsilon_n \boldsymbol{\vartheta}_{n}) \nabla^2_{\textbf{z}_n \textbf{z}_n} \textbf{f}_n \times \textbf{1}_K^{{\scriptsize{\textnormal{T}}}}]_{(\varepsilon_n=0)}\\ &+&o \nonumber
 \end{eqnarray}
From (\ref{utilityrobust2}), the first term in (\ref{tailorseriesmappingf}) is equal to $-\nabla_{\textbf{a}_n}v_n(\textbf{a}_n, \textbf{f}_n)$. Since $\textbf{f}_{n}$ is a linear function of system parameters, the last term in (\ref{tailorseriesmappingf3}) is equal to zero. From (\ref{mappingG}), we have
\begin{eqnarray} \label{perturbedtailorseries}
 \widetilde{\mathcal{F}}_n(\textbf{a})= \mathcal{F}_n(\textbf{a})- \frac{\varepsilon_n}{2}[\nabla^2_{\textbf{a}_n \boldsymbol{f_n}} v_n(\textbf{a}_n, \textbf{f}_n) \times \nabla_{\textbf{z}_n}\boldsymbol{f_n} ]-\frac{\varepsilon^2_n}{3!}[\nabla_{\textbf{a}_n \boldsymbol{f_n}^2}^3 v_n(\textbf{a}_n, \textbf{f}_n) \times (\nabla_{\textbf{z}_n}\boldsymbol{f_n})^2\times \textbf{1}_K^{{\scriptsize{\textnormal{T}}}}] +o,
 \end{eqnarray}
From \textbf{A1}, all the derivatives of $v_n(\textbf{a}_n, \textbf{f}_n)$ are bounded. Therefore, the last terms in (\ref{perturbedtailorseries}) are bounded, and $\widetilde{\mathcal{F}}_n(\textbf{a})$ is the perturbed bounded version of $\mathcal{F}_n(\textbf{a})$.

\section{Proof of Theorem 2}
1) Consider the bounded perturbation of mappings $\mathcal{F}(\textbf{a})$ and $\widetilde{\mathcal{F}}(\textbf{a})$ caused by
variations in system parameters as $\mathcal{Q}= \| \mathcal{F}(\textbf{a})- \widetilde{\mathcal{F}}(\textbf{a})\|_2  \quad \forall \textbf{a}_n \in \widehat{\mathcal{A}}$. Since the strategy space of all users in each dimension is bounded as in (\ref{An}), and the uncertainty region is bounded and convex, this region is also bounded, i.e., $ q^{\text{max}}= \max_{\textbf{a} \in \mathcal{A}} \min_{\widetilde{\textbf{f}}_{n} \in \Re_{n}(\textbf{a}_{-n})} \| \mathcal{F}(\textbf{a})-\widetilde{\mathcal{F}}(\textbf{a})\|_2\leq \infty, \, \quad  \forall n \in \mathcal{N}$.
Any solution to the worst-case robust optimization in
(\ref{utilityrobsut}) corresponds to a realization of
$VI(\mathcal{A},\widetilde{\mathcal{F}})=VI(\mathcal{A},
\mathcal{F}+\textbf{q})$, where $\textbf{q}=q\times (\textbf{1}^{\text{T}}_K)_1^N$, and $ q \in \mathcal{Q}$ depends on $(\tilde{\textbf{a}}_n,\widetilde{\textbf{f}}_{n})$ obtained by (\ref{utilityrobsutsaddlepoint}) for each user, and always $q \leq q^{\text{max}}$. When $\mathcal{F}(\textbf{a})$ is continuous and strictly monotone on the closed convex set $\mathcal{A}$, meaning that $\boldsymbol{\Upsilon}$ is a $P$ matrix, the solution to $VI(\mathcal{A},\mathcal{F}+\textbf{q})$, denoted by $\Phi(\textbf{q})$, is a monotone and single-valued mapping on its domain (Exercise 2.9.17 in \cite{PangVI}), i.e.,
\begin{equation}\label{MONOTONE}
    \forall \textbf{q}_{i}=q_i \times (\textbf{1}^{\text{T}}_K)_1^N, \textbf{q}_{j}=q_j \times (\textbf{1}^{\text{T}}_K)_1^N, \,\, q_i , q_j \in \mathcal{Q} \quad \Longrightarrow (\Phi(\textbf{q}_{i})-\Phi(\textbf{q}_{j}))(\textbf{q}_{i}-\textbf{q}_{j})=0,
\end{equation}
Thus, when $q_{i} \neq q_{j}$, we have $\Phi(\textbf{q}_{i})=\Phi(\textbf{q}_{j})$, which is single valued on $\mathcal{Q}$, i.e., a unique solution for all $q \in Q$. This completes the proof of the uniqueness of RNE under the $P$ property of $\boldsymbol{\Upsilon}$.  2) Recall that when $\boldsymbol{\Upsilon}$ is a $P$ matrix,
$\mathcal{F}(\textbf{a})$ is strictly monotone, and the utility is strictly convex. Since $\mathcal{A}$ is convex in $\mathbb{R}^K$, and $\mathcal{F}(\textbf{a}): K \rightarrow \mathbb{R}^K$ is a continuous mapping on $\mathcal{A}$, the solution to $VI(\mathcal{A},\mathcal{F}+\textbf{q})$ is always a compact and convex set (Corollary 2.6.4 in \cite{PangVI}). Also, since $\textbf{a}_{n}^{*}$ is the optimum value of this convex set for $VI(\mathcal{A},\mathcal{F})$, i.e., $q=0$, any point in this set is less than $\textbf{a}_{n}^{*}$, which is a solution to $VI(\mathcal{A},\mathcal{F}+\textbf{q})$. Note that $\widetilde{\textbf{a}}_{n}^*$
belongs to this set. Since $\boldsymbol{\Upsilon}$ is a $P$ matrix and $\mathcal{F}$ is strictly monotone, we have
\begin{equation}\label{}
 \forall \,\, \textbf{a}_{n}\leq
\textbf{a}_{n}^*  \Longrightarrow u_{n}(\textbf{a}_{n},
\textbf{a}_{-n})\leq
u_{n}(\textbf{a}_{n}^{*},\textbf{a}_{-n}^{*})\quad \forall
\textbf{a} \in \mathcal{A}
\end{equation}
which is also valid for $\widetilde{\textbf{a}}_{n}^*$. As such, the utility at the RNE is less than that at the NE. 3) Since $\mathcal{F}$ is strongly monotone, there is a unique solution denoted by $\widetilde{\textbf{a}}^*=\Phi^*(\textbf{q})$, which can be considered as the worst-case robust solution to $\widetilde{\mathcal{G}}$ for $\|\textbf{q}\|_2\leq \|\boldsymbol{\Delta}\|_2$. Now, both $\textbf{a}_{n}^{*}$ and $\widetilde{\textbf{a}}_{n}^{*}$ must satisfy
\begin{equation}\label{inequality1}
0 \leq (\Phi^*(\textbf{q})-\Phi^*(\textbf{0}))(\mathcal{F}(\Phi^*(\textbf{\textbf{0}}))) \:\qquad \text{and} \:\qquad 0 \leq (\Phi^*(\textbf{0})-\Phi^*(\textbf{q}))(\mathcal{F}(\Phi^*(\textbf{q}))+\textbf{q})
\end{equation}
where $\textbf{0}=(\textbf{0}_K)_1^N$ and $\textbf{0}_K$ is the $K \times 1$ all zero vector. By rearranging \ref{inequality1}, we get
\begin{equation}\label{inequality3}
(\Phi^*(\textbf{0})-\Phi^*(\textbf{q}))(\mathcal{F}(\Phi^*(\textbf{0}))-\mathcal{F}(\Phi^*(\textbf{q})))
\leq (\Phi^*(\textbf{0})-\Phi^*(\textbf{q}))\textbf{q}
\end{equation}
Since $\Phi^*(\textbf{q)}$ is the co-coercive function of $\textbf{q}$ (Proposition 2.3.11 in \cite{PangVI}), the left hand side of (\ref{inequality3}) is always less than $c_{\text{sm}}\|\Phi^*(\textbf{0})-\Phi^*(\textbf{q})\|^{2}$. Using Schwartz inequality for the right hand side, we have
 \begin{equation}\label{inequality4}
\|(\Phi^*(\textbf{0})-\Phi^*(\textbf{q}))\textbf{q}\|_2 \leq  \|\Phi^*(\textbf{0})-\Phi^*(\textbf{q})\|_2 \| \textbf{q} \|_2.
\end{equation}
Since $\Phi^*(\textbf{0})$ and $\Phi^*(\textbf{q})$ correspond to
$\textbf{a}_{n}^*$ and $\widetilde{\textbf{a}}_{n}^*$,
respectively, (\ref{upperbound of variations}) can be obtained.  4) Since the difference between utility functions of each user at RNE and at NE is equal to first term of the Taylor series of $u^k_n(\textbf{a}_{n}, \textbf{f}_{n})$ with respect to all variations in the strategies of user $n$ and other users, we have
\begin{eqnarray}\label{}
  u^k_n(a_{n}^k, f_{n}^k)-v^k_n(a_{n}^k, f_{n}^k)\approx \varepsilon_n ( \frac{\partial v^k_n (a_{n}^k, f_{n}^k)}{ \partial a^k_n} \times \frac{\partial a_{n}^k}{\partial \varepsilon_n}+\frac{\partial v^k_n (a_{n}^k, f_{n}^k)}{ \partial f^k_n} \times \frac{\partial f^k_n}{\partial \varepsilon_n} )\quad \forall n \in \mathcal{N}, \quad k \in \mathcal{K},
\end{eqnarray}
which is equivalent to
\begin{eqnarray}\label{Theoream42}
  u^k_n(a_{n}^k, f_{n}^k)-v^k_n(a_{n}^k, f_{n}^k)&\approx&\varepsilon_n (\frac{\partial v^k_n (a_{n}^k, f_{n}^k)}{ \partial a^k_n} \times \frac{\partial a_{n}^k}{\partial \varepsilon_n}+ \frac{\partial v^k_n (a_{n}^k, f_{n}^k)}{ \partial f^k_n} \times \frac{\partial f^k_n}{\partial \textbf{a}_{-n}}\frac{\partial \textbf{a}_{-n}}{ \partial \varepsilon_n})  \nonumber \\
  &=& \varepsilon_n (\frac{\partial v^k_n (a_{n}^k, f_{n}^k)}{ \partial a^k_n} \times \frac{\partial a_{n}^k}{\partial \varepsilon_n}+ \frac{\partial v^k_n (a_{n}^k, f_{n}^k)}{ \partial f^k_n} \times \sum_{m\neq n} x_{nm}^k \frac{\partial a_{m}^k}{ \partial \varepsilon_n})
\end{eqnarray}
When $\varepsilon_n$ is sufficiently small, the derivative of the  strategy of each user is approximately equal to \begin{equation}\label{approaximation}
 \lim_{\varepsilon \rightarrow 0}  \frac{\tilde{a}_n^{*k}-a_n^{*k}}{\varepsilon_n}= \frac{\partial a_{n}^k}{ \partial \varepsilon_n},
\end{equation}
By expanding (\ref{Theoream42}) for all users in the game,  we have
\begin{equation}\label{diffrencebetweenustilituesappecndix}
    \| v(\textbf{a}^*)-u(\tilde{\textbf{a}}^*) \|_2 \approx \| \textbf{W}(\textbf{a}^*)\| _2 \times \|\textbf{a}^*-\widetilde{\textbf{a}}^*\| ,
\end{equation}
by replacing (\ref{upperbound of variations}) into (\ref{diffrencebetweenustilituesappecndix}), the approximation (\ref{diffrencebetweenustilitues}) is obtained.
\section{Proof of Proposition 1}
1) From (\ref{utilitylogconvex}), the best-response of the NACG is
\begin{equation}\label{Proofpro11}
a_{n}^{k}=[\lambda_n^{\frac{1}{\theta}}-\frac{c_{n}^{k}+y_{n}^k}{x_{nn}^k}-\sum_{m\neq n} \frac{x_{nm}^{k}}{x_{nn}^{k}} a_{m}^{k}]_{a_{nk}^{\text{min}}}^{a_{nk}^{\text{max}}} ,
\end{equation}
where the Lagrange multiplier $\lambda_n$ for each user is so chosen to satisfy the sum constraint $\sum_{k=1}^{K}a_{n}^{k}= a_{n}^{\texttt{max}}$. Therefore, the best response of this problem can be written as an \textit{AVI}, denoted by $AVI(\mathcal{A}, \mathcal{M})$, where $\mathcal{M}_n$ is obtained from (\ref{AVIlog}) \cite{PangVI}. 2) For this case, the game has a unique NE when $\mathcal{M} (\textbf{a})$ is strongly monotone or when $\textbf{M}(\textbf{a})\triangleq (\textbf{M}_{nm}(\textbf{a}))_{m,n=1}^{N}$ is positive definite (Sections 2.3 and 2.4 in \cite{PangVI}). By some rearrangements, we have
\begin{equation}\label{Proofpro12}
    \textbf{M}(\textbf{a})=(\textbf{M}(k)(\textbf{a}))_{k=1}^{K}
\end{equation}
where $[\textbf{M}(k)]_{nm}=\frac{x^{k}_{nm}}{x^{k}_{nn}}$. When all $\textbf{M}(k)$ are positive definite matrices, $\mathcal{M}(\textbf{a})$ is strongly monotone. Consider $\textbf{M}^{\mathrm{max}}$ in (\ref{mmax}), and its strictly lower and upper triangle matrix as $\textbf{M}_{\text{Low}}^{\mathrm{max}}$, and $\textbf{M}_{\text{upp}}^{\mathrm{max}}$, respectively. Let $\textbf{b}=\textbf{I}-\textbf{M}_{\text{Low}}^{\mathrm{max}} - \textbf{M}_{\text{upp}}^{\mathrm{max}}$ and $p_n=\|\textbf{a}_n\|_2$. When for any $\textbf{p}=[p_1,\cdots p_N]>0$, we have
\begin{equation}\label{Proofpro14}
 \max_{n \in \mathcal{N}}p_n \sum_{n=1}^{N} B_{nm}p_m>0 , \quad
\end{equation}
all $\textbf{M}(k)$ for all $k$ are positive definite matrices (Proposition 1 in \cite{Luoiterativewaterfilling}). By rearranging (\ref{Proofpro14}), we obtain (\ref{conditionproposition1}). Therefore, when (\ref{Proofpro14}) holds, \textit{AVI} has a unique solution and consequently, the NE is unique.

\section{Proof of Theorem 3}
1) From Lemma 1, the map of RACG is the perturbed map of NACG. Since the map of NACG is linear for the utility function (\ref{utilitylogconvex}), the perturbed map is
\begin{equation}\label{mappingM3}
    \widetilde{\mathcal{M}}^k(\textbf{a}) =w^k_n+ \sum_{m=1}^{N}\frac{\widetilde{x}_{nm}^k}{x_{nn}^k}a_{m}^k , \quad \forall \widetilde{x}_{nm}^k \in \mathcal{R}_n^k
\end{equation}
where $\widetilde{\mathcal{M}}^k(\textbf{a})$ and $w^k_n$ are the $k^{\text{th}}$ elements of $\widetilde{\mathcal{M}}(\textbf{a})$ and $\textbf{w}_n$, respectively. Now, (\ref{mappingM3}) can be rewritten as
\begin{equation}\label{mappingM}
     \widetilde{\mathcal{M}}^k(\textbf{a}) =w^k_n+ \sum_{m=1}^{N}(\frac{\widetilde{x}_{nm}^k}{x_{nn}^k} +\frac{\widetilde{x}_{nm}^k-x_{nm}^k}{x_{nn}^k})a_m^k \leq w^k_n+ \sum_{m=1}^{N}\frac{x_{nm}^k}{x_{nn}^k}a_m^k+\epsilon_{n}^k \|\textbf{a}_{-n}^k\|,
\end{equation}
Therefore, the map at the RNE is (\ref{AVIlog}). 2) Since $a_{m}^{k}$ is bounded in $[a_{mk}^{\text{min}},a_{mk}^{\text{max}}]$ and the uncertainty region is bounded, the value of $\boldsymbol{\epsilon}_{n}^{k}\|\textbf{a}^{k}_{-n}\|_2$ is bounded. Therefore, for any bounded uncertainty region considered by users in the RACG, its \textit{AVI} can be rewritten as $AVI=(\mathcal{A},\mathcal{M}+\textbf{m})$, where $\textbf{m}=(\textbf{m}_{n})_{n=1}^{N}=(\textbf{w}_{n}+\widehat{\mathcal{M}}_n)_{n=1}^{N}$ and $\|\textbf{m}\|_2< \infty $. Therefore, the RNE is the perturbed solution to $AVI=(\mathcal{A},\mathcal{M})$. From Theorem 4.3.2 in \cite{PangVI}, when $\textbf{M}$ is semicopositive, the \textit{AVI} has a unique solution for any value of $\textbf{m}$. Therefore, the robust game has a unique solution for any bound on the uncertainty region.  3) The second part can be obtained the same as in part 2 of Theorem 1. Recall that $\mathcal{M}$ is strongly monotone on
$\mathcal{A}$ when there exist a $c_\text{sm}$, such that for all
$\textbf{a}^1=(\textbf{a}_n)_{n \in \mathcal{N}}$, and
$\textbf{a}^2=(\textbf{a}_n)_{n \in \mathcal{N}}$ we have
\begin{equation}\label{strongmonotone}
    (\textbf{a}^1-\textbf{a}^2)(\mathcal{M}(\textbf{a}^1)-\mathcal{M}(\textbf{a}^2))\geq c_\text{sm}\parallel \textbf{a}^1-\textbf{a}^2 \parallel.
\end{equation}
When $e_{n}^{k}=(a_{n}^{k1}-a_{n}^{k2})$, for our linear
formulation we have
\begin{eqnarray}
    \lefteqn{(\textbf{a}_n^1-\textbf{a}_n^2)(\mathcal{M}_n(\textbf{a}^1)-\mathcal{M}_n(\textbf{a}^2))=(\textbf{a}_n^1-\textbf{a}_n^2)(\sum_{m=1}^{N}(\textbf{M}_{nm}\textbf{a}_m^1-\textbf{M}_{nm}\textbf{a}_m^2))=}  \nonumber \\   & &  \sum_{k=1}^{K}(a_{n}^{k1}-a_{n}^{k2})[\sum_{m=1}^{N}M^{kk}_{nm}(a_{n}^{k1}-a_{n}^{k2}) \geq  \sum_{k=1}^{K} (e_{n}^{k})^2-\sum_{m=1, \, m \neq n}^{N}
|\sum_{k=1}^{K}e_{m}^{k}\frac{x_{nm}^{k}}{x_{mm}^{k}}e_{n}^{k}| \nonumber\\
&& \geq \sum_{k=1}^{K} (e_{n}^{k})^2- \sum^{N}_{m=1, \, m \neq
n}(\sum_{k=1}^{K}e_{n}^{k})^2 \max_{k \in
K}\frac{x_{nm}^{k}}{x_{mm}^{k}}(\sum_{k=1}^{K}(e_{m}^{k})^2)^2 \geq \| \textbf{e}_{n}\|_2
\sum_{m=1}^{N}[\textbf{M}^{\text{max}}]_{nm}\|\textbf{e}_{m}\|_2. \nonumber
\label{lastinequality}
\end{eqnarray}
Therefore, for all $n \in \mathcal{N}$ we have
\begin{equation}\label{1}
  (\textbf{a}^1-\textbf{a}^2)(\mathcal{M}(\textbf{a}^1)-\mathcal{M}(\textbf{a}^2))\geq  \textbf{e}\textbf{M}^{\text{max}}\textbf{e}
\end{equation}
Given (\ref{1}) and summing over $q$, since $\textbf{M}^{\text{max}}$ is positive semi-definite, we have
\begin{equation}\label{endoftheoram3}
 \textbf{e}\textbf{M}^{\text{max}}\textbf{e}\geq  \lambda_\text{min}(\textbf{M}^{\text{max}}) \|\textbf{e}\|_2
\end{equation}
Therefore, the lower bound of the strong monotonicity constant in $\mathcal{M}(\textbf{a})$ is obtained. By replacing (\ref{endoftheoram3}) into (\ref{upperbound of variations}), part 3 of Theorem 3 can be obtained.

\section{Proof of Theorem 4}
When the solution to (\ref{proximalresponsemap}) is obtained by the non-expansive or contraction mapping, the distributed algorithm for the proximal response map converges. For any vector $\textbf{z} \in \widehat{\mathcal{A}}$ in (\ref{proximalresponsemap}), we have
\begin{eqnarray} \label{Theorem 41}
  (\textbf{z}- \widehat{\textbf{a}}(\textbf{b}_1)) [\widehat{\mathcal{F}}(\widehat{\textbf{a}}(\textbf{b}_1), \textbf{b}_1)+ \widehat{\textbf{a}}(\textbf{b}_1)-\textbf{b}_1]\geq 0\\
 \label{Theorem 42} (\textbf{z}- \widehat{\textbf{a}}(\textbf{b}_2)) [\widehat{\mathcal{F}}(\widehat{\textbf{a}}(\textbf{b}_2), \textbf{b}_2)+ \widehat{\textbf{a}}(\textbf{b}_2)-\textbf{b}_2]\geq 0
  \end{eqnarray}
Considering $\textbf{z}=\widehat{\textbf{a}}(\textbf{b}_2)$ in (\ref{Theorem 41}) and $\widehat{\textbf{z}}=\textbf{a}(\textbf{b}_1)$ in (\ref{Theorem 42}), from the above two inequalities we get
\begin{eqnarray} \label{Theorem 43}
 0\leq (\widehat{\textbf{a}}(\textbf{b}_2)- \widehat{\textbf{a}}(\textbf{b}^1)) [\widehat{\mathcal{F}}(\widehat{\textbf{a}}(\textbf{b}^1), \textbf{b}^1)+ \widehat{\textbf{a}}(\textbf{b}^1)-\textbf{b}^1]+ (\widehat{\textbf{a}}(\textbf{b}^1)- \widehat{\textbf{a}}(\textbf{b}^2)) [\widehat{\mathcal{F}}(\widehat{\textbf{a}}(\textbf{b}^2), \textbf{b}^2)+ \widehat{\textbf{a}}(\textbf{b}^2)-\textbf{b}^2]\\ \label{Theorem 44}
 =(\widehat{\textbf{a}}(\textbf{b}^2)- \widehat{\textbf{a}}(\textbf{b}^1))[\widehat{\mathcal{F}}(\widehat{\textbf{a}}(\textbf{b}^1), \textbf{b}^1)-\widehat{\mathcal{F}}(\widehat{\textbf{a}}(\textbf{b}^2), \textbf{b}^2)]  -\|\widehat{\textbf{a}}(\textbf{b}^2)- \widehat{\textbf{a}}(\textbf{b}^1)\|+ (\widehat{\textbf{a}}(\textbf{b}^2)- \widehat{\textbf{a}}(\textbf{b}^1))^T (\textbf{b}^1-\textbf{b}^2)
\end{eqnarray}
Recall that $\widehat{\mathcal{F}}_n= -\nabla_{\textbf{a}_n} u_n(\textbf{a}_n, \textbf{f}_n+ \varepsilon_n \boldsymbol{\vartheta}_{n}) - \varepsilon_n \nabla_{\widetilde{\textbf{f}}_n}  u_n(\textbf{a}_n, \textbf{f}_n+ \varepsilon_n \boldsymbol{\vartheta}_{n})\times \nabla_{\textbf{a}_n} \boldsymbol{\vartheta}_{n}$, $\nabla_{\textbf{a}_n} \widehat{\mathcal{F}}_n= - \nabla^2_{\textbf{a}_n,\textbf{a}_n} u_{n}+\varepsilon_1 \times \nabla_{\textbf{a}_n \textbf{a}_n\textbf{f}_n}^3 u_n$ and $\nabla_{\textbf{a}_m} \widehat{\mathcal{F}}_n= -\nabla_{\textbf{a}_n \textbf{a}_m}^2 u_{n}+\varepsilon_1 \times \nabla_{\partial \textbf{a}_n \partial^2 \textbf{f}_n}^3 u_n \textbf{x}_{nm}$. When $\frac{\partial^3 v_n }{\partial \textbf{a}_n \partial^2 \textbf{f}_n}=\frac{\partial^3 v_n }{\partial^2 \textbf{a}_n \partial \textbf{f}_n}=0$,
(\ref{Theorem 44}) can be written as
\begin{eqnarray} \label{Theoream467}
&&(\widehat{\textbf{a}}(\textbf{b}^2)- \widehat{\textbf{a}}(\textbf{b}^1))[ \sum_{n \in \mathcal{N}}-\nabla_{\textbf{a}_n \textbf{a}_n}^2 u_{n}](\widehat{\textbf{a}}(\textbf{b}^2)- \widehat{\textbf{a}}(\textbf{b}^1))^{{\scriptsize{\textnormal{T}}}} \\+ && (\widehat{\textbf{a}}(\textbf{b}^1)- \widehat{\textbf{a}}(\textbf{b}^2))[ \sum_{m \in \mathcal{N}, m \neq n}-\nabla_{\textbf{a}_n \textbf{a}_m}^2 u_{n}] (\textbf{b}^1-\textbf{b}^2)^{{\scriptsize{\textnormal{T}}}} \nonumber -\|\widehat{\textbf{a}}(\textbf{b}^2)- \widehat{\textbf{a}}(\textbf{b}^1)\|+ (\textbf{a}(\textbf{b}^2)- \textbf{a}(\textbf{b}^1)) (\textbf{b}^1-\textbf{b}^2)^{{\scriptsize{\textnormal{T}}}} \geq 0 \nonumber
\end{eqnarray}
Let $\widetilde{\alpha}_{n}(\textbf{a})\triangleq$ be the smallest eigenvalue of $-\nabla^{2}_{\textbf{a}_{n}}u_{n}(\textbf{a}_{n},\textbf{f}_n(\textbf{a}_{-n},\textbf{x}_n))$, $\widetilde{\beta}_{nm}(\textbf{a})\triangleq\|-\nabla_{\textbf{a}_{n}\textbf{a}_{m}}u_{n}(\textbf{a}_{n},\textbf{f}_n(\textbf{a}_{-n},\textbf{x}_n))\|$ for $n\neq m$, and $\textbf{z}\triangleq \tau(\textbf{a}^1(\textbf{b}^1),\textbf{b}^1)+(1-\tau)(\textbf{a}^2(\textbf{b}^2),\textbf{b}^2)$. From (\ref{Theoream467}) we get
\begin{equation}\label{theoream48}
    (1+\widetilde{\alpha}_n(\textbf{z}))\|\widehat{\textbf{a}}(\textbf{b}^2)-\widehat{ \textbf{a}}(\textbf{b}^1)\|\leq \sum_{n=1}^{N} \widetilde{\beta}_{nm}(\textbf{z}^n)\|\textbf{b}_{-n}^1-\textbf{b}_{-n}^2\|,
\end{equation}
On the other hand,
 \begin{equation}\label{90}
  - \nabla_{\textbf{a}_n \textbf{a}_n}^2 u_{n} = - \nabla_{\textbf{a}_n \textbf{a}_n} u_n(\textbf{a}_n, \textbf{f}_n- \varepsilon_n \boldsymbol{\vartheta}_{n}) + \varepsilon_n \nabla_{\widetilde{\textbf{f}}_n}  u_n(\textbf{a}_n, \textbf{f}_n- \varepsilon_n \boldsymbol{\vartheta}_{n})\times \nabla_{\textbf{a}_n} \boldsymbol{\vartheta}_{n}.
 \end{equation}
Since the utility is a convex function with respect to $\textbf{a}_n$, the first term in (\ref{90}) is positive and $ \| \nabla_{\textbf{a}_n \textbf{a}_n}^2 u_{n}\| \geq \| \nabla_{\textbf{a}_n \textbf{a}_n}^2 v_{n}\|$. Besides,  \begin{equation}\label{}
 \nabla_{\textbf{a}_n \textbf{a}_m}^2 u_{n} =  \nabla_{\textbf{a}_n \textbf{a}_m} u_n(\textbf{a}_n, \textbf{f}_n+ \varepsilon_n \boldsymbol{\vartheta}_{n}) -\varepsilon_n  \nabla_{\widetilde{\textbf{f}}_n}  u_n(\textbf{a}_n, \textbf{f}_n- \varepsilon_n \boldsymbol{\vartheta}_{n})\times \nabla_{\textbf{a}_n} \boldsymbol{\vartheta}_{n}, \end{equation}
which leads to $\|\nabla_{\textbf{a}_n \textbf{a}_m}^2 u_{n}\|\leq \| \nabla_{\textbf{a}_n \textbf{a}_m}^2 v_{n}\|$. From these two inequalities, (\ref{theoream48}) can be written as
\begin{equation}\label{theoream49}
    (1+\alpha_n(\textbf{z}))\|\widehat{\textbf{a}}(\textbf{b}^2)- \widehat{\textbf{a}}(\textbf{b}^1)\|\leq \sum_{n=1}^{N} \beta_{nm}(\textbf{z}^n)\|\textbf{b}_{-n}^1-\textbf{b}_{-n}^2\|,
\end{equation}
For (\ref{theoream49}), when $\boldsymbol{\Upsilon}$ is a $P$ matrix, the proximal response map in (\ref{proximalresponsemapn}) is a contraction mapping (Proposition 12.17 in \cite{Palomar2010}), and converges to a unique RNE from any initial point.

\section{Proof of Proposition 2}
Variation in utility function of each user for variation in the bound of uncertainty region is
\begin{eqnarray}\label{utilityRNE}
    \lim_{\boldsymbol{\varepsilon} \rightarrow 0} \nabla_{\boldsymbol{\varepsilon}}u_n(\textbf{a}_n, \widetilde{\textbf{f}}_n)= (\nabla_{\textbf{a}_n} v_n(\textbf{a}_n, \textbf{f}_n) \times \textbf{1}_K)  \times \nabla_{\varepsilon_n} \textbf{a}_n +(\nabla_{\textbf{f}_n} v_n(\textbf{a}_n, \textbf{f}_n)  \textbf{x}_{nm} )\nabla_{ \varepsilon_m}\textbf{a}_m  \nonumber\\+\varepsilon_n \times \nabla^2_{\textbf{a}_n \textbf{f}_n} v_n(\textbf{a}_n, \textbf{f}_n) \times \textbf{X}_{nm} \times \nabla_{ \varepsilon_n}\textbf{a}_n + \varepsilon_n \textbf{X}_{nm} \nabla^2_{\textbf{f}_n \textbf{f}_n} v_n(\textbf{a}_n, \textbf{f}_n)  \times \nabla_{\varepsilon_m} \textbf{a}_m,
\end{eqnarray}
where $\textbf{X}_{nm}$ is the $K \times K$ matrix, $[\textbf{X}_{nm}]_{kk}=(x_{nm}^k)^2$,  $\boldsymbol{\varepsilon}=[\varepsilon_1,\cdots,\varepsilon_N]$, and $\nabla$ is the column gradient vector.  The last two terms in (\ref{utilityRNE}) are always positive, because of \textbf{A3} and \textbf{A2}. The first term in (\ref{utilityRNE}) is always negative because $v_n(\textbf{a}_n, \textbf{f}_n)$ is increasing according to $\textbf{a}_n$ and $ \nabla_{\varepsilon_n} \textbf{a}_n<0$. The second term in (\ref{utilityRNE}) is always positive because $v_n(\textbf{a}_n, \textbf{f}_n)$ is a decreasing function of  $\textbf{f}_n$ and $\nabla_{\varepsilon_n}  \textbf{a}_m<0$. Therefore, the social utility increases when the negative terms of (\ref{utilityRNE}) are less than the positive terms. By some rearrangement and matrix manipulation, the condition for negative semi-definiteness of $\textbf{W}$ can be obtained.

\section{Negative Semi-Definiteness Condition for Affine \textit{VI}}
1) Consider $AVI(\mathcal{B}, \mathcal{N}+\textbf{n})$, where $\mathcal{B}$ is a closed convex set, $\mathcal{N}(\textbf{a})$ is the monotone map related to $\mathcal{B}$, $\textbf{a} \in \mathcal{B}$, and $\textbf{n} \in \mathcal{B}$ is the vector with bounded positive values. Let $\Phi(\textbf{n})$ be the solution to $AVI(\mathcal{B}, \mathcal{N}+\textbf{n})$. When $\mathcal{N}(\textbf{a})$ is strongly monotone, $\Phi(\textbf{n})$ is monotone (Corollary 2.9.17 in [31]). When $\Phi(\textbf{n})$ is a monotone and decreasing function, we have
\begin{eqnarray}
  (\Phi(\textbf{n})-\Phi(\textbf{0}))(\mathcal{N}(\Phi(\textbf{0}))>0  \quad \quad \textbf{n} \in \mathcal{B}  \quad \forall \textbf{n}>0 \label{H1}\\
    (\Phi(\textbf{0})-\Phi(\textbf{n}))(\mathcal{N}(\Phi(\textbf{n})+\textbf{n})>0  \quad \quad \textbf{n} \in \mathcal{B} \quad \forall  \textbf{n}>0\label{H2}
\end{eqnarray}
subtracting (\ref{H1}) from (\ref{H2}), we get
\begin{eqnarray}
      (\Phi(\textbf{0})-\Phi(\textbf{n}))(\mathcal{N}(\Phi(\textbf{n})-\mathcal{N}(\Phi(\textbf{0}))+\textbf{n}(\Phi(\textbf{0})-\Phi(\textbf{n}))>0 \label{H3}
\end{eqnarray}
The above inequality leads to $\mathcal{N}(\Phi(\textbf{n}))>\mathcal{N}(\Phi(\textbf{0}))$. Because of the affinity in \textit{AVI}, (\ref{H3}) is
\begin{equation}\label{H4}
    \mathcal{N}\textbf{f}>0 \quad \forall \textbf{f} \in \mathcal{B}
\end{equation}
where $\textbf{f}=\Phi(\textbf{n})-\Phi(\textbf{0})$ is a negative vector and $\textbf{f} \in \mathcal{B}$. Since $\mathcal{B}$ is a convex and closed region, we have
\begin{equation}\label{H4}
    \textbf{f}^*\mathcal{N} \textbf{f} < 0, \quad \quad \forall \textbf{f} \in \mathcal{B}
\end{equation}
which is equivalent to the semi-negative matrix definition. 2) From above, when the strategy of each user in the power control game in (\ref{propostion1AVI}) is a decreasing function, $\mathcal{M}$ is a semi-negative matrix. Therefore, the social utility increases when the strategy of each user is reduced. Obviously, $\mathcal{M}$ is semi negative when $-\mathcal{M}$ is semi-positive, which leads to (\ref{conditionproposition1forpowercontrolgame}).
\end{appendices}

\bibliographystyle{IEEEtran}

\begin{figure}
\centering
\includegraphics [height=5cm,width=9.5cm] {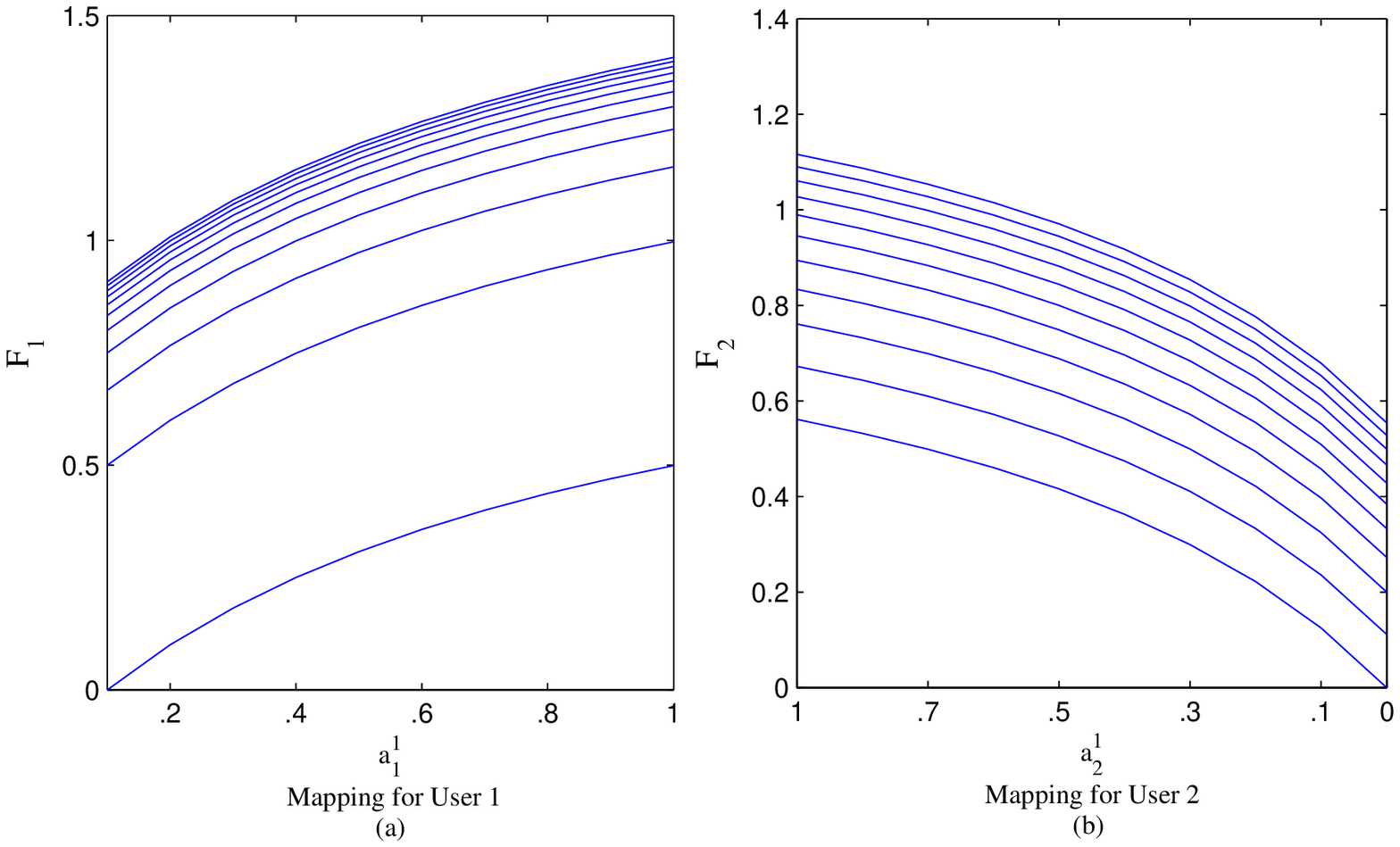}
\caption{Monotone mapping $\mathcal{F}_1$ and
$\mathcal{F}_2$ for the power control game when $\boldsymbol{\Upsilon}$ is a $P$ matrix.}{\label{1}}
\end{figure}
\small \small
\begin{table}[ht]
\caption{Examples of ACG} \label{tableexample}
\centering 
\vspace{-0.2 in}
\begin{tabular}{c c c c c} 
\hline\hline 
Example & $v_{n}^{k}(a_n^k,f_{n}^{k})$ & $f_{n}^{k}$ & $x_{nm}^k$ &$y_{n}^k$ \\ [0.5ex] 
\hline 
1 &  $\log(1+\frac{a_{n}^{k}}{f_{n}^{k}})$ & $\sum_{m \in \mathcal{N},~m \neq n}\bar{h}_{nm}^{k}a_{m}^{k}+\bar{\sigma}_{n}^{k}$ & $\bar{h}_{nm}^{k}$ & $\bar{\sigma}_{n}^{k}$ \\ 
2 & $\mu_{n}^{k}-\nu_{nn}^{k}a_{n}^{k}-f_{n}^{k}$ & $\sum_{m \in \mathcal{N},~m \neq n} \nu_{nm}^{k} a_{m}^{k}$ & $\nu_{nm}^{k}$  & $0$ \\
 [1ex] 
\hline 
\end{tabular}
\end{table}
\small \small
\begin{table*}[h]
\caption{Distributed Algorithm}\label{distributedalgorith}
\centering
\vspace{-0.2 in}
\begin{tabular}{l}
\hline \hline
\textbf{Distributed Algorithm for Proximal Response Map}\\
\hline \hline
\textbf{Inputs for Each User}
\\ $\mathcal{T}=[1,\cdots,T]$: Users' iterations, $\varepsilon_n$: Uncertainty region for user $n$,
and  $0<\zeta<< 1$: Interrupt criteria for all users,
\\ \textbf{Initialization} For $t=0$,
 \\ set a feasible strategy $\textbf{a}_n(0) $ and a random value $\textbf{f}_n(0)$ for all $n \in \mathcal{N}$,
\\ \textbf{Iterative Algorithm} \\For $t=1,\cdots,T$ and $T \rightarrow \infty$ %
\\Update the transmit strategy $\textbf{a}_n(t)= \max_{\textbf{a}_n \in \mathcal{A}_n} \Psi_n- \frac{1}{2} \|\textbf{a}_n(t)-\textbf{a}_n(t-1)\|_2^2$ for all users,
\\Each user transmits the value of $\textbf{a}_n(t)$ to other users, and
\\ measures the aggregate effects of other users $\textbf{f}_n(t)$,
\\ If $\|\textbf{a}_n(t-1)-\textbf{a}_n(t)\|\leq \zeta $ End, otherwise $t=t+1$, continue;
\\ \hline \hline
\end{tabular}
\end{table*}
\small \small
\begin{table}[h]
\caption{Opportunistic Algorithm for Increasing Social Utility} \label{tableopportunesticalgorithm}
\centering
\vspace{-0.2 in}
\begin{tabular}{l}
\hline \hline
\textbf{First Stage:} All users play $\mathcal{G}$ to reach the NE with utility $v^*_{n}$.
\\ \textbf{Second Stage:} If $\alpha_{n}^{\text{min}} < \sum_{m\neq n} \beta_{m}^{\text{max}}$, $\forall n \in
\mathcal{N}$
\\ \textbf{Initialization:} Let $\widetilde{u}(0)=\sum_{n=1}^{N}v^*_{n}$,~$0<\chi<1$,~ $\omega(0)=0$, and~$0<\delta<< 1$ 
\\ \textbf{Iterative Algorithm}\\ For $t_1=1,\cdots,T_1$;
\\1) Consider uncertainty region $\varepsilon_n=\omega(t_1)$, where $\omega(t_1)=t_1 \times \chi  $for all n;
\\2) $ \textbf{a}_n(t_1)= \max_{\textbf{a}_n \in \mathcal{A}_n} \Psi_n-\frac{1}{2} \|\textbf{a}_n(t_1)-\textbf{a}_n(t_1-1)\|_2^2$ for all user;
\\3) User $n$ transmits with $\textbf{a}_n(t_1)$, measures $\textbf{f}_{n}(t_1)$, and calculates $\widetilde{u}_n(t-1)$;
\\4) Users exchange their $\textbf{a}_n$ and $\widetilde{u}_n(t_1)$,
\\5) The social utility in $t_1$ is calculated via $\widetilde{u}(t_1)=\sum_{n=1}^{N} \widetilde{u}_n(t_1)$;
\\6) If $\widetilde{u}(t_1)>\widetilde{u}(t_1-1)$, and $\| \widetilde{u}(t_1)-\widetilde{u}(t_1-1)\|> \delta $: Continue; Otherwise: End. \\[1ex]
\hline \hline
\end{tabular}
\end{table}
\end{document}